\documentclass[aps,pre]{revtex4}
\usepackage{graphicx}

\def\lapl{\nabla^2}
\def\grad{\nabla}
\def\degree{^\circ}
\newcommand{\vct}[1]{\mathbf{#1}}

\begin{document}
\title{Pearling instability of nanoscale fluid flow confined to a chemical
channel}
\author{J.\ Koplik}
\affiliation{Benjamin Levich Institute and Department of Physics,
City College of the City University of New York, New York, NY 10031}
\author{T.\ S.\ Lo}
\affiliation{Benjamin Levich Institute and Department of Physics,
City College of the City University of New York, New York, NY 10031}
\author{M.\ Rauscher}
\affiliation{Max-Planck-Institut f\"{u}r Metallforschung, Heisenbergstr.\ 3, 
70569 Stuttgart, Germany, and \\
Institut f{\"u}r Theoretische und Angewandte Physik, Universit\"{a}t Stuttgart, Pfaffenwaldring 57, 70569 Stuttgart, Germany}
\author{S.\ Dietrich}
\affiliation{Max-Planck-Institut f\"{u}r Metallforschung, Heisenbergstr.\ 3, 
70569 Stuttgart, Germany, and \\
Institut f{\"u}r Theoretische und Angewandte Physik, Universit\"{a}t Stuttgart, Pfaffenwaldring 57, 70569 Stuttgart, Germany}

\date{\today}

\begin{abstract}
We investigate the flow of a nano-scale incompressible ridge of
low-volatility liquid along a ``chemical channel'': a long,
straight, and completely wetting stripe embedded in a planar substrate,
and sandwiched between two extended less wetting solid regions.  Molecular 
dynamics simulations, a simple long-wavelength approximation, and a full 
stability analysis based on the Stokes equations are used, and give 
qualitatively consistent results. While thin
liquid ridges are stable 
both statically and during flow, a (linear) pearling instability develops 
if the thickness of the ridge exceeds half of the width of the channel. 
In the flowing case periodic bulges propagate along the channel and 
subsequently merge due to nonlinear effects. However, the ridge does 
not break up even when the flow is unstable, and the qualitative behavior 
is unchanged even when the fluid can spill over onto a partially wetting 
exterior solid region.
\end{abstract}
\maketitle

\section{Introduction}
\label{sec:intro}

In recent years substantial efforts have been invested in miniaturizing 
chemical processes by building microfluidic systems. The ``lab on a 
chip'' concept integrates a great variety of chemical and physical 
processes into a single device \cite{giordano01,mitchell01,stone01} in a 
similar way as an integrated circuit incorporates many electronic devices 
into a single chip. These microfluidic devices do not only allow for cheap 
mass production but they can operate with much smaller quantities of 
reactants and reaction products than standard laboratory equipments. Even 
though most available microfluidic devices today have micron sized 
channels, further miniaturization is leading towards the nano-scale.

There are two main lines of development for microfluidic systems. The 
first one encompasses systems with closed channels, essentially 
microfabricated tubes. However, closed channel systems have the 
disadvantage that they can be easily clogged by solute particles such as 
colloids or large bio-polymers.

The second type consists of systems which are open with a free
liquid-vapor interface, where the fluid is confined laterally not by
geometrical but by chemical walls.  The idea is 
that the liquid will be guided by lyophilic stripes on an otherwise 
lyophobic substrate \cite{darhuber01,gau99,dietrich05}. The substrate 
surfaces can be structured chemically by printing or photographic 
techniques. All these techniques are confined to two dimensions.

While liquid flow in closed channel systems can be pumped by applying a 
pressure difference between the inlet and the outlet, the liquid-vapor 
interface of an open system would yield to the pressure (like a very soft 
rubber tube) and droplets would form.  Furthermore, upon pumping flow directions in
interconnected open channel systems would be
difficult to predict because small droplets may or may not inflate into larger
ones, depending on details of the channel geometry and the filling state.
However, there are a number of alternative ways to drive a 
liquid through a chemical channel. The most obvious one is to use
inertia to drive the flow, e.g., centrifugal forces or gravity, as occurs
when a liquid film flows down an inclined 
plane. For films with thicknesses in the nanometer range one needs stronger 
accelerations in order to achieve a decent throughput. These can be realized 
most conveniently on a rotating disc (e.g., a spin coater or a 
centrifuge). Electric forces are also commonly used in order to drive 
flows in small channels, e.g., by using the electro-osmotic effect
which, however, only works for liquids containing ions.
Since water is the most 
interesting liquid for biological applications, this is not a 
serious restriction. Depending on the wall potential, a charged layer 
a few {\AA} in thickness forms and an electric field applied parallel
to the substrate will set the liquid in this layer in motion. While the 
electro-osmotic effect drives the liquid molecules at the substrate, the 
Marangoni effect drives the molecules at the liquid-vapor interface. The 
origin of the Marangoni forces are gradients in surface tension which 
arise from temperature gradients or from gradients in surfactant 
concentrations. The fourth method to cause flow in a chemical channel
is wicking, i.e., the sucking of a droplet into a chemical channel by
capillary forces.

Here we consider inertial driving of the liquid, which is one of 
the most promising methods to generate flow in open microfluidic devices. 
Marangoni forces could be induced easily by local heating of the liquid,
but for volatile fluids like water this will lead to enhanced and 
undesirable evaporation. Wicking is slow for longer channels and 
ceases to act as a driving force
once the channel is filled. In principle, electrical forcing 
has the advantage that the direction of the force can be controlled on 
very small scales, but the required fields are high and might lead to 
electrolysis. In contrast, inertial driving can be realized easily by 
centrifugal forces and does not affect the liquid sample.

\begin{figure}
\begin{center}
\includegraphics[width=0.4\linewidth]{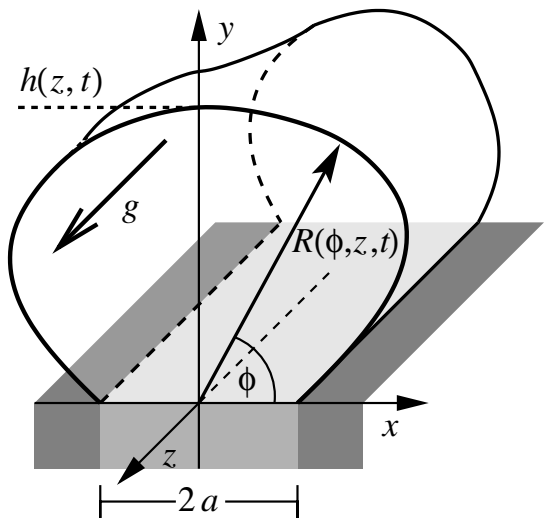}
\end{center}
\caption{\label{fig:wurst} 
A liquid ridge on a chemical wetting stripe of width $2a$ embedded in a
solid substrate, which is non-wetting.  Gravity or inertial forces act on
the liquid in the direction parallel to the $z$-axis. The liquid-vapor
interface is parameterized in cylindrical coordinates as $r=R(\phi,z,t)$,
and the local height $h(z,t)=R(\pi/2,z,t)$\/.}
\end{figure}

Our focus in this paper is the stability and robustness of one-dimensional 
channel flow (i.e., flow in a straight channel) with inertial forcing 
aligned with the channel direction. The flow geometry is sketched in 
Fig.~\ref{fig:wurst}. In the absence of forcing, a 
chemical channel filled homogeneously by a non-volatile liquid of
fixed volume is
unstable with respect to droplet formation when the contact angle with
the triple-line pinned at the channel edge
exceeds 90$\degree$ \cite{davis80,brinkmann05}\/. This pearling instability is 
surface tension driven and similar to the Rayleigh-Plateau instability. 
The resulting drop shapes have been studied extensively in Ref.\
\cite{brinkmann02}. Of particular interest is the question of spilling 
onto the lyophobic substrate, which can lead to cross-talk between
neighboring
channels. In the force-free case this happens for large droplets on 
chemical stripes on a partially wetting embedding substrate.

In Sec.~\ref{sec:md} we describe the formulation of the molecular dynamics 
(MD) simulations for flow on a chemical channel and discuss the special 
case of unidirectional flow quantitatively. In Sec.~\ref{sec:pearl} 
we discuss the appearance of pearling instabilities in channel flows and, 
in particular, we study the influence of the liquid ridge height
(effectively, the channel
depth) and of the wetting properties of the surrounding
substrate on the stability of
the liquid ridge.  The full 
linear stability analysis of a Stokes flow in Sec.~\ref{sec:stab} as well 
as the hydrodynamic long-wavelength approximation developed in 
Sec.~\ref{sec:lw} are in qualitative agreement with the MD simulations. 
We discuss the results in Sec.~\ref{sec:disc}.

\section{Molecular Dynamics Simulations}
\label{sec:md}

The MD simulations carried out here employ standard techniques 
\cite{at87,fs02} and a 
simple molecular model for a non-polar liquid in contact with a 
thermally-fluctuating solid \cite{kb95,kb00}, based on atoms interacting 
via an adjustable Lennard-Jones potential
\begin{equation}
V_{\rm LJ}(r) = 4\,\epsilon\, \left[ \left( {r\over\sigma} \right)^{-12} -
c_{ij}\, \left( {r\over\sigma} \right)^{-6}\ \right].
\label{eq:lj}
\end{equation}
The interaction is cut off at $r_c=2.5\,\sigma$, and shifted by a linear 
term so that the force vanishes smoothly there. In this section we 
employ so-called ``MD units'' derived from this potential, with the atomic 
core size 
$\sigma$ as the unit of length, the potential well depth $\epsilon$ as the 
unit of energy, and $\tau\equiv \sigma(m/\epsilon)^{1/2}$ as the unit of 
time $t$, where $m$ is the mass of the liquid atoms. We simulate a 
generic non-polar liquid rather than any particular laboratory material,
but typical numerical values are $\sigma\approx 0.3$~nm,
$\epsilon/k_B \approx 140$K, and $\tau\approx 2$~ps\/.
A monatomic Lennard-Jones fluid exhibits a high vapor pressure and a 
very broad liquid-vapor interface. These are inconvenient properties in
the present context and are ameliorated by using a FENE potential
\begin{equation} 
V_F(r)=-\frac{1}{2}\, k_F\, r_0^2\, \ln\left(1-{r^2\over r_0^2}\right) 
\end{equation}
to group the atoms into freely-jointed chain molecules which are four 
atoms in length.  The numerical values of the constants in $V_F$ are
$k_F=30\epsilon/\sigma^2$ and $r_0=1.5\sigma$, following
Ref.~\cite{grest86}\/. This leads to a significant reduction of the
vapor pressure and thus yields a liquid of low volatility.
The solid atoms are tethered to two layers of fcc
lattice sites $\{\vct{r}_{0i}\}$ with lattice constant $1.55\,\sigma$  
using a stiff spring
\begin{equation}
V_S(\vct{r}_i)=\frac{1}{2}\,k_S\, \left(\vct{r}_i-\vct{r}_{0i}\right)^2
\qquad i=1,\ldots,n_S
\end{equation}
with spring constant $k_S=100\,\epsilon/\sigma^2$, allowing the liquid and 
solid to exchange energy and momentum. The substrate surface has a
(100) crystallographic orientation.
The solid atom mass is chosen as $100\,m$, so as to have approximately the 
same oscillation frequency as the LJ interactions and a common time step 
in the numerical integration. During the simulations, a Nos\'e-Hoover 
thermostat maintains a constant temperature $1.0\,\epsilon/k_B$\/.
Under these conditions, the fluid condenses into a liquid of number
density $0.79\,\sigma^{-3}$, in equilibrium  with a modest amount of vapor.

The coefficient $c_{ij}$ in Eq.~(\ref{eq:lj}) is used to vary the strength 
of the attractive interaction between atomic species $i$ and $j$. The 
intra-liquid coefficient has the (standard) value $c_{LL}=1.0$\/. 
With the other modeling choices and parameter values given above, the
liquid wets the solid completely, partially, and not at all for the choices 
$c_{LS}=$ 1.0, 0.75, and 0.0, respectively.  The corresponding contact 
angles for a drop on a homogeneous solid with these values of $c_{LS}$ are 
0$\degree$, approximately 90$\degree$, and 180$\degree$, respectively.
In 
the latter case and in the absence of gravity, a drop actually drifts off
the surface.  The surface tension of the liquid is found from a standard 
simulation of a slab of liquid coexisting with vapor to be $\gamma = 
0.46\,\epsilon/\sigma^2$, and the liquid viscosity is obtained from a 
separate simulation of Couette flow as $\eta = 3.60\,m/(\sigma\,\tau)$\/.

\begin{figure}
\includegraphics[width=0.8\linewidth]{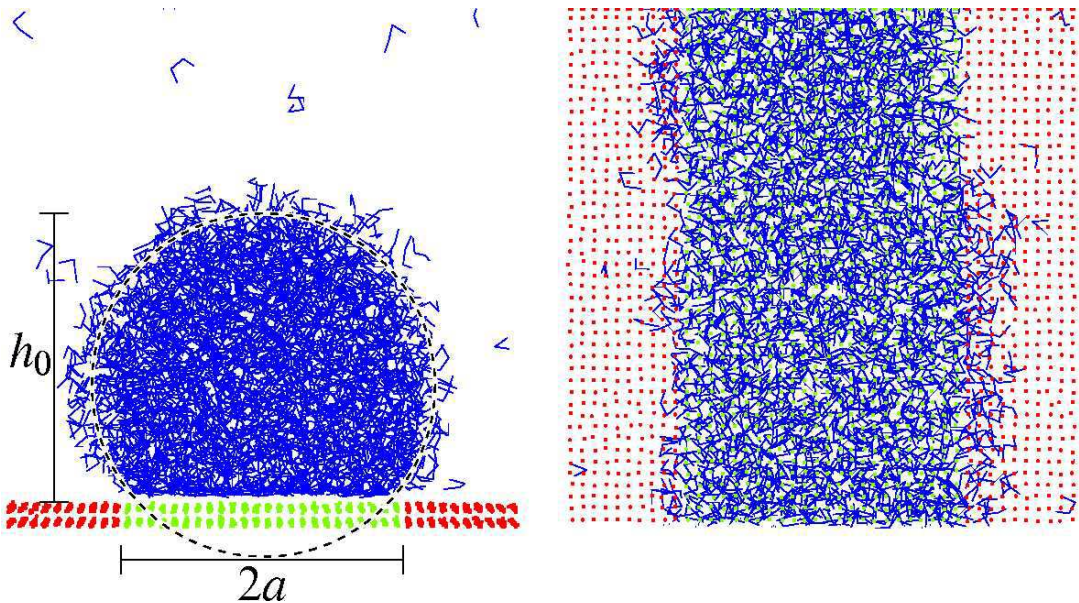}
\caption{\label{fig:snap} Snapshot of a typical initial molecular
configuration.  The left and right views are along and top down on the
stripe, respectively. The liquid molecules are depicted as the three
``bonds'' joining the four atoms, while substrate atoms with wetting
($c_{LS}=1$) and non-wetting ($c_{LS}=0$) properties are light green and
dark red dots, respectively. The average spacing between solid atoms is
0.78$\sigma$\/. The liquid ridge has a circular cross section (as
indicated by the dashed circle) and a height $h_0$ (at which the density
attains half its value in the center), which depends on the liquid volume
and the stripe width $2\,a$\/.}
\end{figure}

A snapshot of the basic simulated configuration is shown in 
Fig.~\ref{fig:snap}, i.e., the individual atoms in a liquid ridge on a 
wetting stripe for a case where the exterior of the channel is completely 
non-wetting. The left figure shows an end-on view (parallel to the stripe) 
and the right figure a section of the top view.
Note that the interaction does confine the liquid to the wetting region, 
and that the interface strongly fluctuates. 
In this and all other 
simulations, we apply periodic boundary conditions in the plane of the 
substrate in order to avoid end effects. To construct the atomic 
configurations, we place atoms on the sites of a cylindrical section of a 
fcc lattice and allow them to move under the action of the LJ and FENE 
potentials given above until the atomic positions become disordered and the 
internal energy stabilizes. This configuration does not correspond to
global thermodynamic equilibrium because, as we shall see, a uniformly
shaped liquid ridge is hydrodynamically unstable; however, this
configuration can be described in terms of a local equilibrium.
The resulting cylinder of liquid is translated to lie just 
above and parallel to the wetting stripe of a solid substrate which has 
been independently allowed to reach local equilibrium, so that the liquid 
is attracted towards the substrate and adjusts itself to form an
irregular circular ridge with height $h_0$ as shown in
Fig.~\ref{fig:snap}\/.  During the course of the equilibration process,
lasting typically several hundred $\tau$, the liquid surface fluctuates at
a 5\% level in an apparently random manner. Systematic changes in the
shape of the ridge occur only on longer time scales.

\begin{figure}
\includegraphics[width=0.4\linewidth]{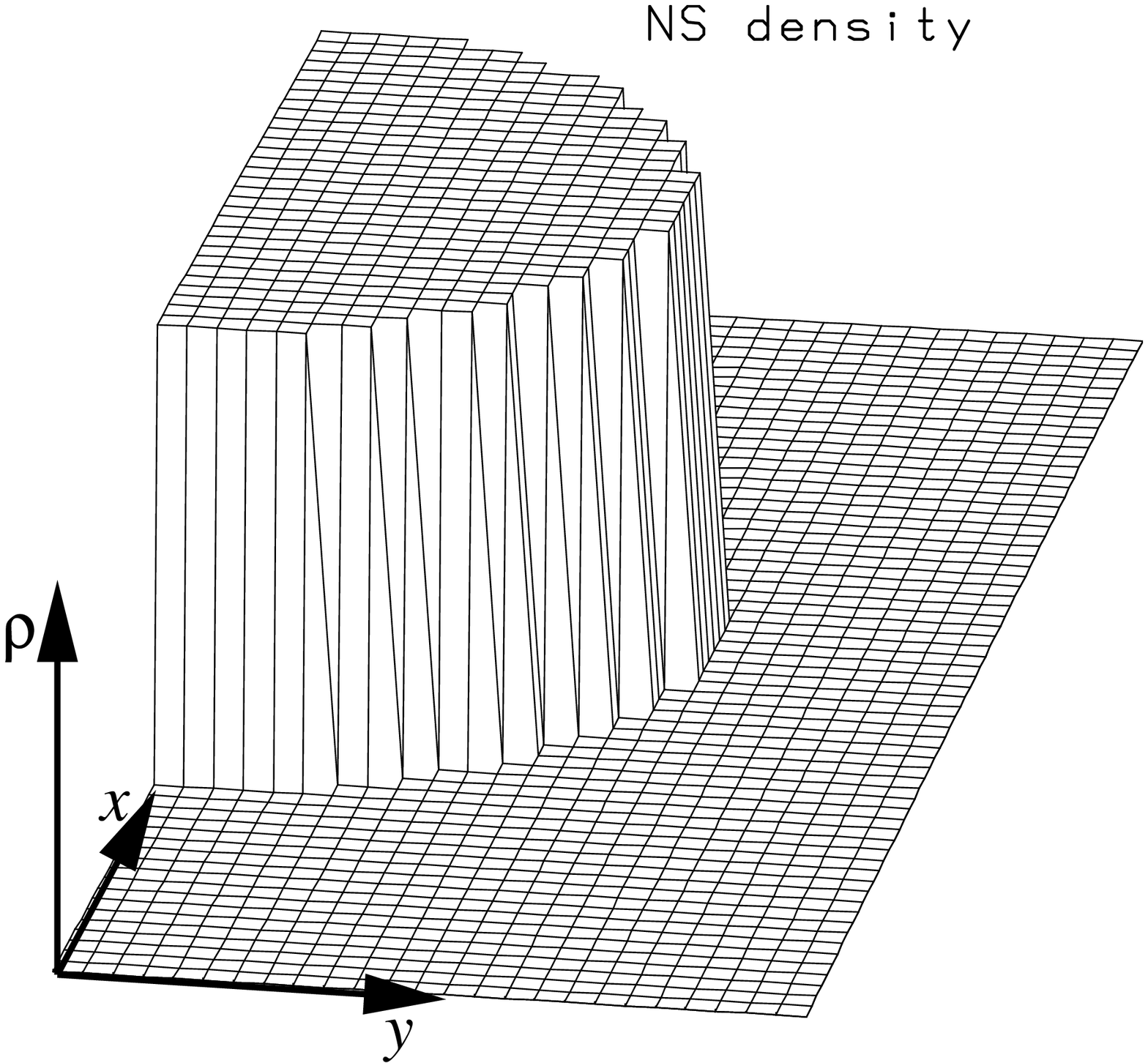}
\includegraphics[width=0.4\linewidth]{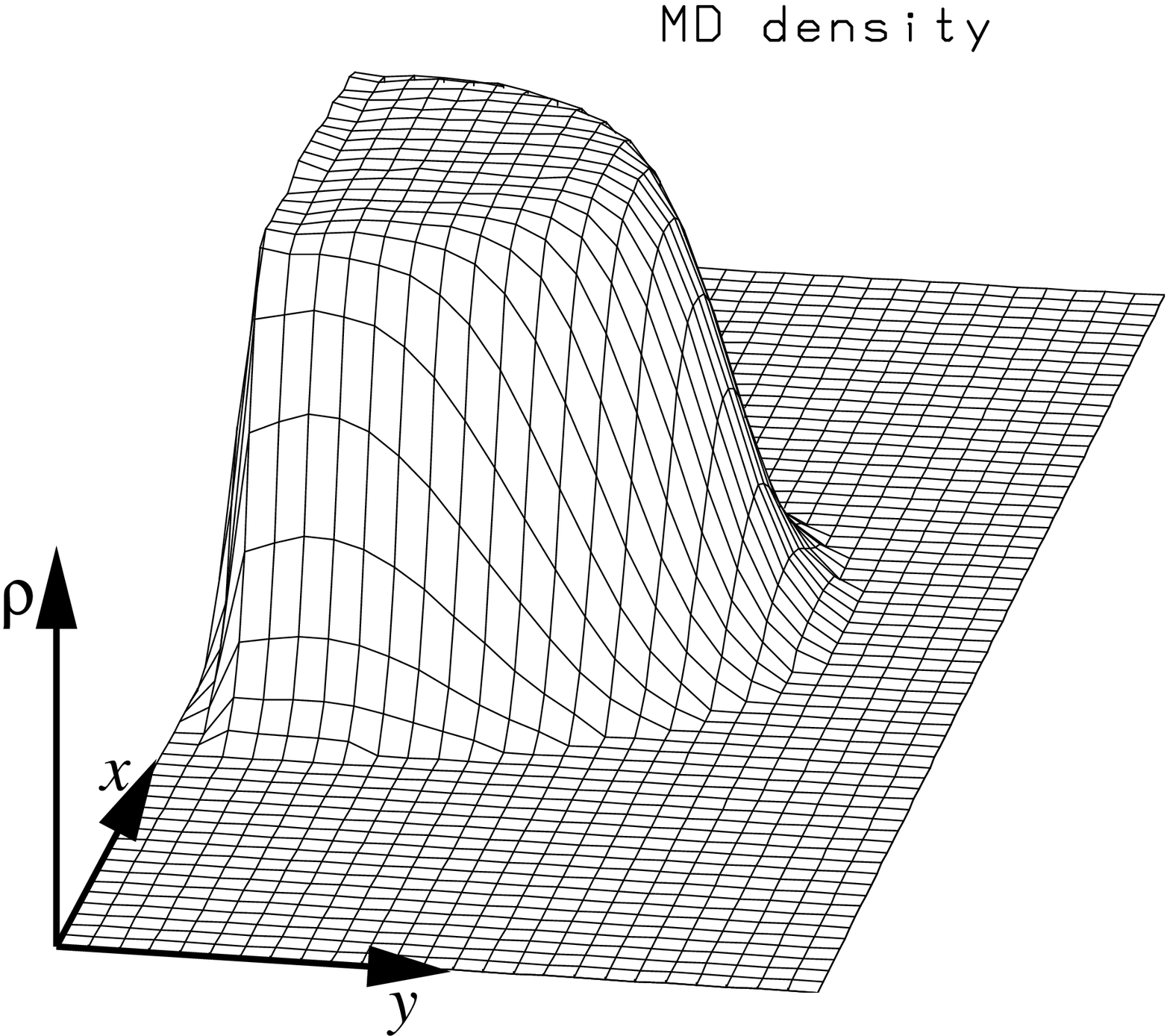} \\

\vspace*{0.4in}

\includegraphics[width=0.4\linewidth]{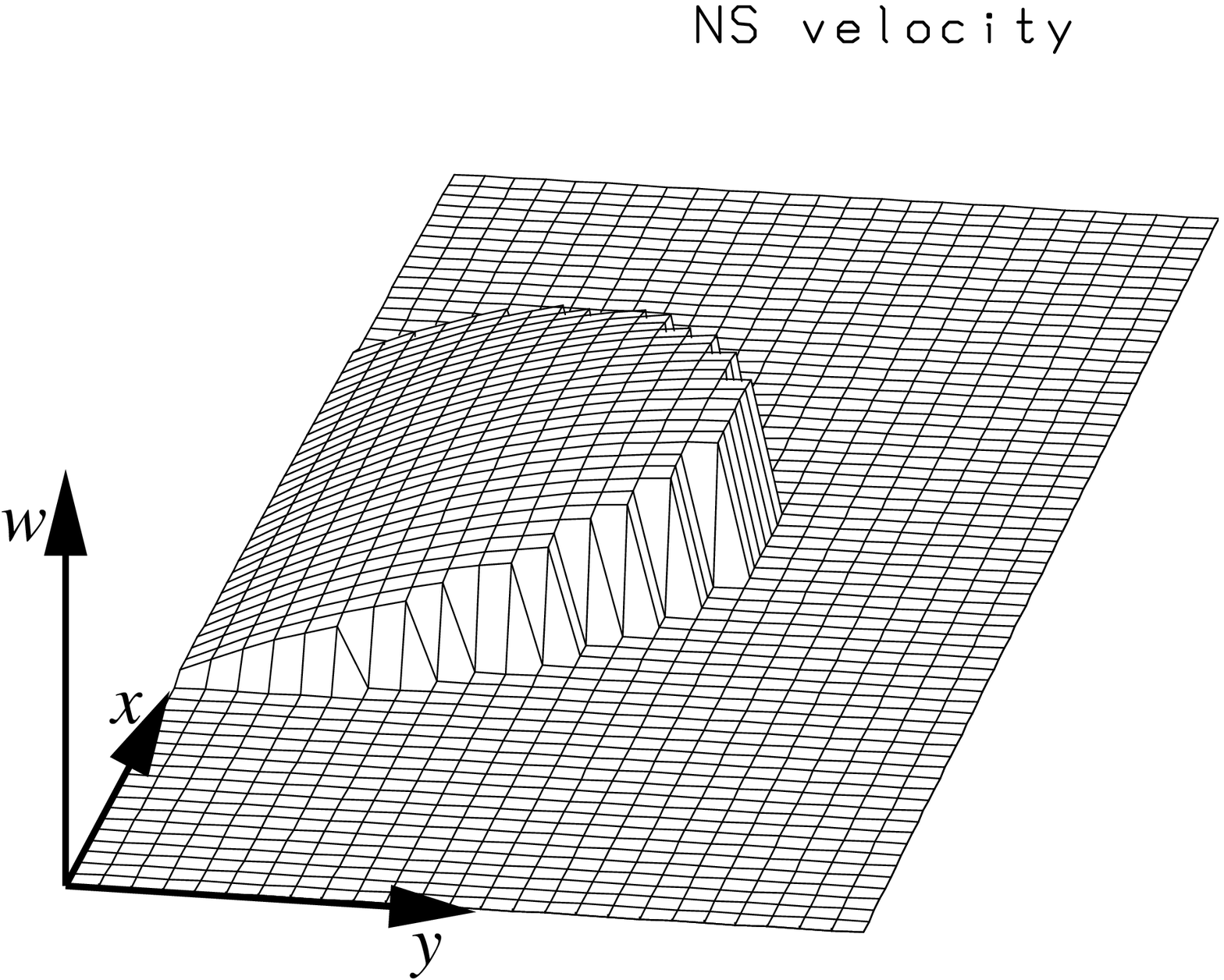}
\includegraphics[width=0.4\linewidth]{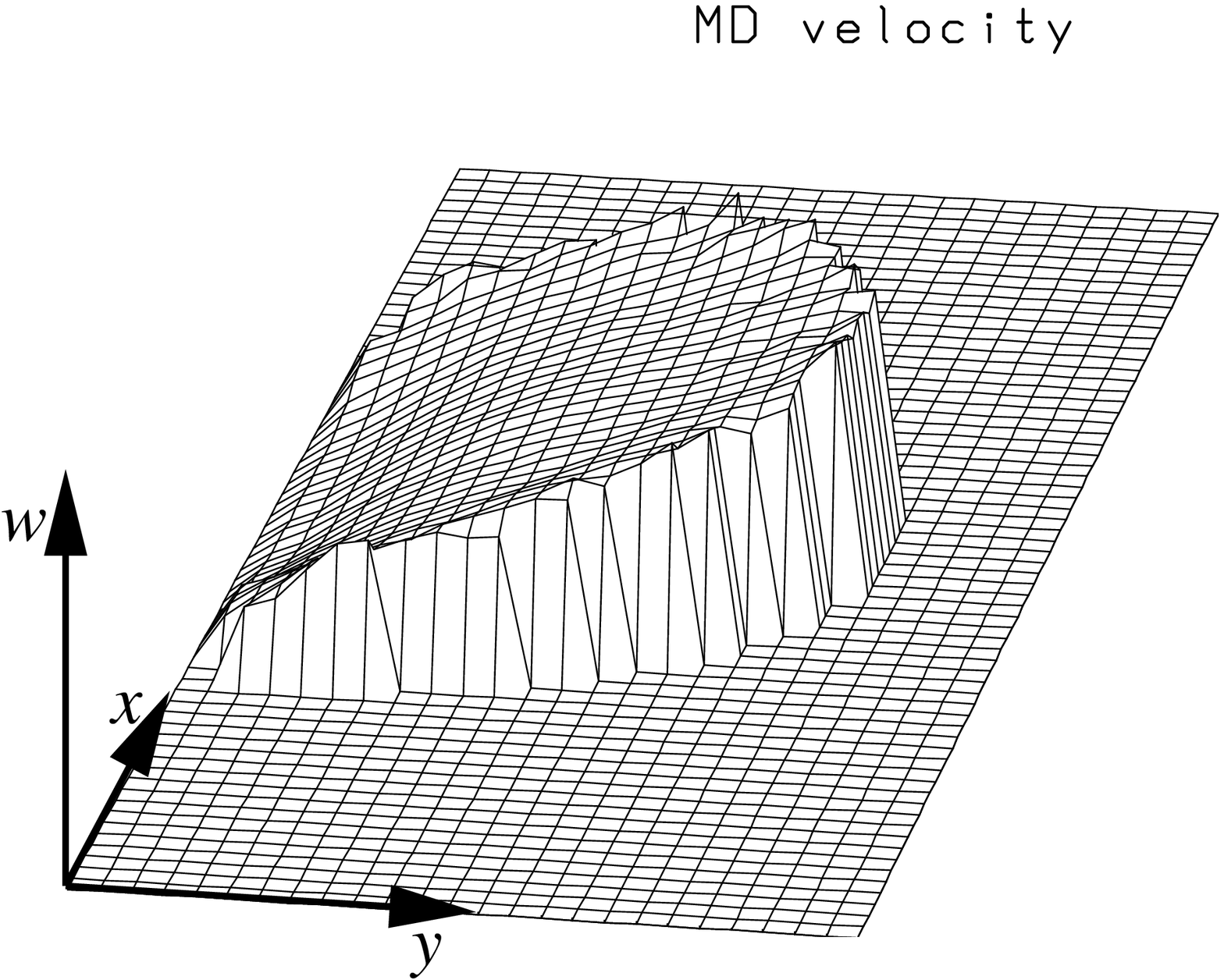} \\

\vspace*{0.4in}

\includegraphics[width=0.4\linewidth]{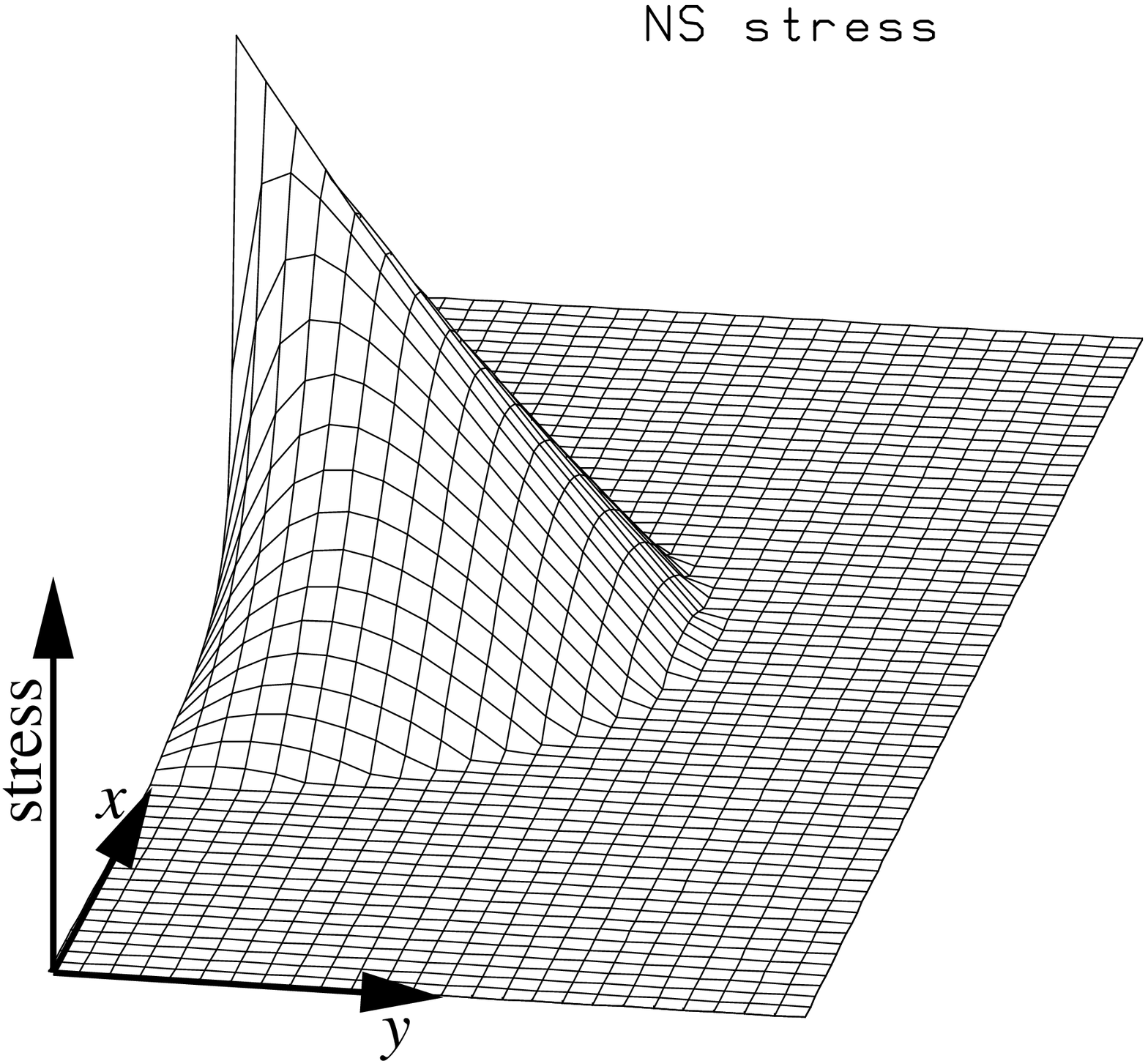}
\includegraphics[width=0.4\linewidth]{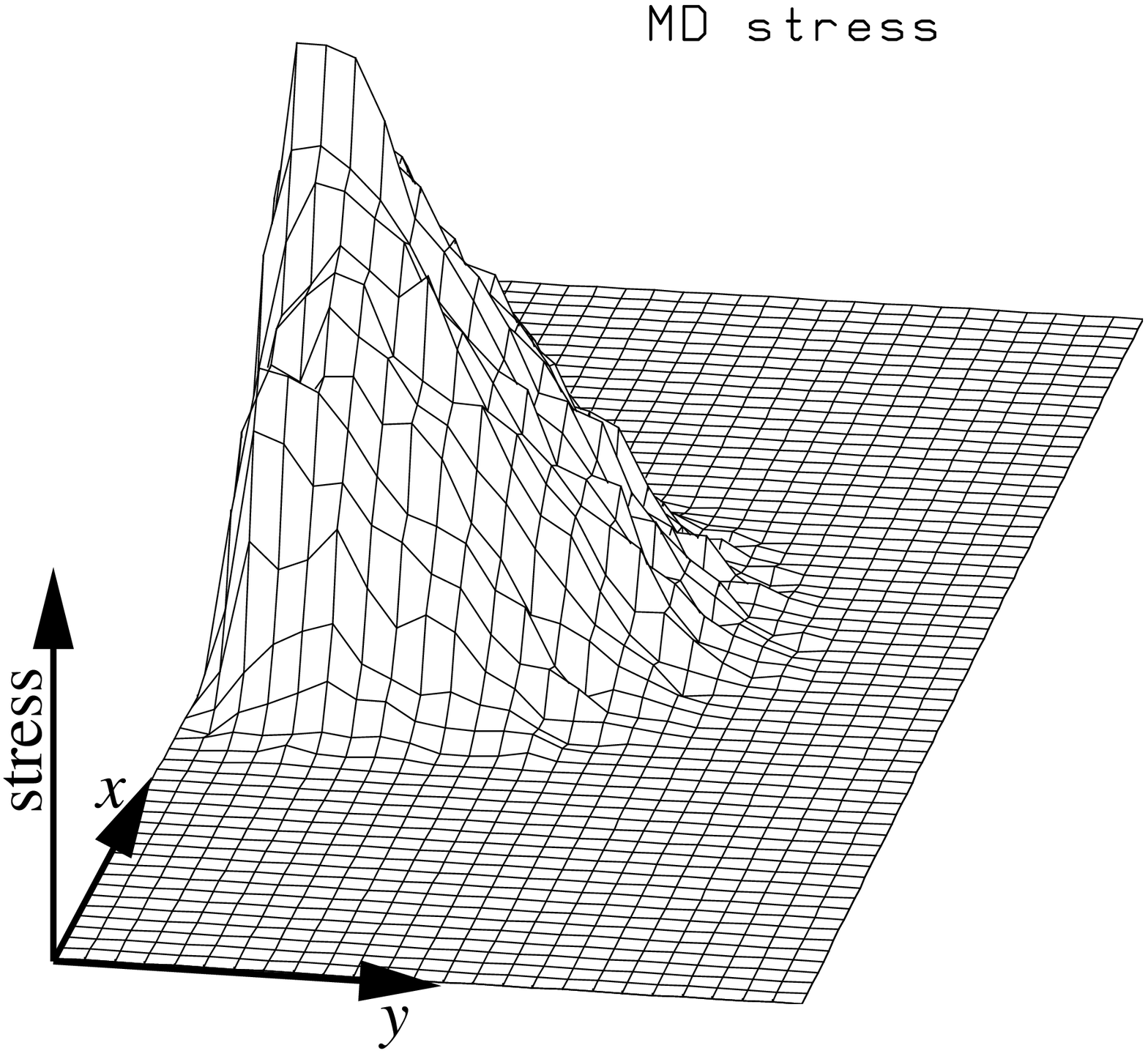} \\
\caption{\label{fig:flow_comp} 
\baselineskip 14pt
Comparison of the density, axial velocity
and $r$-$z$ shear stress fields in a liquid ridge with initial height
$h_0=a=16\sigma$ for unidirectional flow with $g=0.01$  (see
Fig.~\protect\ref{fig:wurst} for the coordinate definitions)\/. In each
frame of the figure the base is the $x$-$y$ plane, with the substrate (not
shown) to the left.  The height of the surface represents the value of the
field depicted there, and the grid spacing is $\sigma$\/. In the density
plots, the plateau height, representing the central liquid density, is
$0.79\,\sigma^{-3}$\/. In the Navier-Stokes (NS) velocity and stress plots,
the maximum values are $0.225\,\sigma/\tau$ and $0.107\,m/(\sigma\tau^2)$,
respectively, and for each field the same scales and units are used for NS
and MD.  The MD fields are set to zero in bins (of size
$\sigma\times\sigma$) where the density is below $0.01\,\sigma^{-3}$\/.
The exterior of the substrate is nonwetting ($c_{LS}=0$) whereas the
stripe is wetting ($c_{LS}=1$)\/.} 
\end{figure}

\normalsize

In order to simulate liquid flow, we apply a uniform force 
$m\,g\,{\vct{e}_z}$ to each liquid atom, where the axis of the stripe is 
the $z$-direction (see Fig.~\ref{fig:wurst})\/. The results presented in this
paper involve values $g=0.01$--$0.05\,\sigma/\tau^2$\/, and the corresponding 
Reynolds numbers, based on the 
observed mean flow velocities and the height of the liquid ridge, are
$\text{Re}=0.1$--$10.0$\/. Smaller values of $g$ are perhaps more
realistic but require longer computations for the flow and for any changes
in interfacial shape to develop. In any case, these low-Reynolds number
simulations are in the linear regime where phenomena simply scale in time
with $1/g$\/.  At higher $g$, the liquid is often driven off the substrate.

In order to relate the MD simulation results to macroscopic hydrodynamics, 
we consider the special case in which the liquid occupies just half of a 
cylinder with a 90$\degree$ contact angle, with initial height $h_0=a$. 
As discussed below in 
Sects.~\ref{sec:stab} {and \ref{sec:lw}}, this case is neutrally stable 
with respect to changes in interface shape and the shape of the ridge
does not change significantly over the time of the simulation. Furthermore,
this case may be directly compared 
(over moderate time intervals) with solutions of the Navier-Stokes (NS) 
equations for unidirectional flow parallel to the stripe axis.  Furthermore, 
the unidirectional NS flow field may be 
computed in an almost closed form by a Fourier series method (see 
Sec.~{\ref{sec:stab}} for details). In Fig.~\ref{fig:flow_comp} we 
show the liquid density, the axial velocity field, and the $r$-$z$ component 
of the shear stress,
as obtained from the analytic solution and from the MD simulations, at the 
same resolution. (The explicit analytic form of the shear stress can be
obtained from the formulas in
Sec.~\ref{sec:stab} and the MD counterpart is the Irving-Kirkwood expression
\cite{at87}.)  
In order to obtain these fields from the simulations, the flow domain in the 
$x$-$y$ plane was divided into square bins of size $\sigma\times\sigma$, 
and the
appropriate quantity in each bin was averaged both over time ($2000\,\tau$) 
and the length of the system in $z$-direction ($517\,\sigma$). 
A larger averaging time interval was impractical due to the eventual 
development of systematic interfacial fluctuations, while smaller sampling 
bins lead to excessive noise in the stress (and wall-induced oscillations in 
the liquid density near the solid \cite{kb95}).  The jaggedness in the NS
figures results from the relatively small number of bins ($31\times 16$)
occupied by the liquid, while that in the MD figures incorporates thermal
fluctuations as well.

The first point to note is that the MD liquid-vapor interface is 
rather diffuse, corresponding to a width of about 7 sampling bins. 
This feature is not easily accounted for in the NS calculation, so that
close agreement should not be expected.  Secondly, we see that while the
shear stress values are in qualitative agreement both in shape and magnitude
and the velocity fields have the same shape, 
the velocity values differ by almost a factor of 2.  This
comparison depends critically on the value chosen for the liquid ridge height
in the NS calculations: a larger height increases the velocity values and 
decreases the stress, so that a different choice could make both agree
to 50\%\/. Here, in computing the NS flow field, we selected the numerical
value of the height $h_0$ based on the reasonable but arbitrary 
definition as the point where the MD density dropped to half 
the value at the center of the ridge (see Figs.~\ref{fig:wurst} and
\ref{fig:snap})\/. The fact that the MD velocities are larger
may result from the fact the free surface region of the ridge exhibits a
transition between liquid and gaseous phases and time-averaged velocities 
in the latter are larger.  In any case, the degree of agreement between MD 
and NS hydrodynamics is worse than in similar comparisons where the liquid
is completely confined 
by walls \cite{kb95}, which we attribute to the diffuse top boundary of the
liquid.  A final remark is that the (thermal) fluctuations in the MD results 
are largest for the stress, moderate for the velocity, and almost invisible
in the density.  The reason is that the stress computation involves the
intermolecular force, which is a rapidly varying function of
atomic separation, while the velocities result from an integration over the
force, which provides some smoothing. The density follows from the atomic
positions, which are a further integral of the velocity, and is thus still
smoother.

\section{Pearling Instabilities}
\label{sec:pearl}

Turning now to the systematic study of pearling instabilities, we have 
conducted a number of simulations in which we
explore the effects of several operating parameters. For a fixed
number of liquid atoms (128,000, or 32,000 molecules) we vary
\begin{itemize}
\item filling: the number of liquid particles per stripe length on the
wetting stripe. In practice, we 
choose different widths for the solid wetting stripe, and since the liquid 
is effectively pinned to the edges of the stripe, it adjusts its height and 
contact angle such that a circular cross section emerges.
\item wetting: the region exterior to the completely wetting stripe can be 
either completely non-wetting (i.e., drying) or partially wetting. 
In the former case the 
liquid is confined to the stripe and does not contact the exterior, but 
in the latter case it can spill over the edges.
\item forcing: the value of the acceleration $g$.
\item length: simulations have been carried out using one-half and
one-quarter of the stripe length and, correspondingly, of the number of
atoms, but with identical cross-sectional shapes. This allows one to
examine the wavelength selection if pearling develops.
\end{itemize}

\begin{figure}
\includegraphics[width=\linewidth]{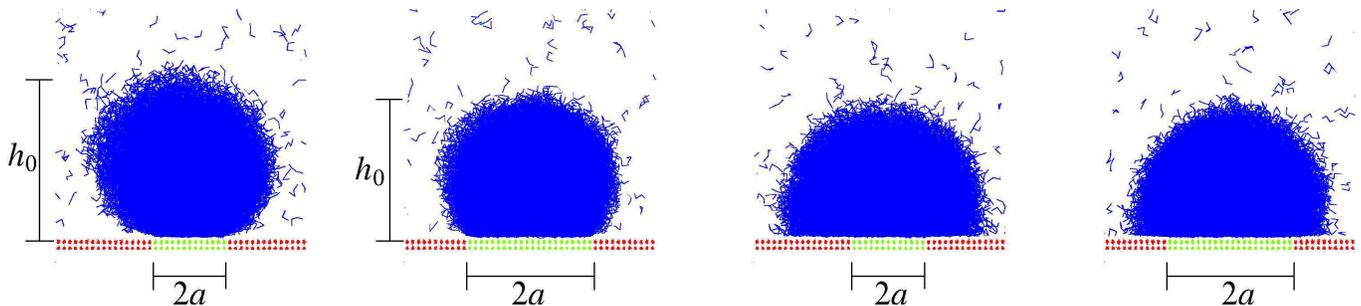}
\caption{\label{fig:cases} Head on view along the $z$-axis of the 
initial configuration of the cases simulated. From left to right: 
non-wetting ($c_{LS}=0$) exterior and narrow wetting ($c_{LS}=1$) stripe
with ${h_0/a=4.0}$; non-wetting exterior and wide wetting stripe with
${h_0/a=2.2}$; partially wetting ($c_{LS}=0.75$) exterior and narrow
wetting stripe with ${h_0/a=3.1}$; partially wetting exterior and wide
wetting stripe with ${h_0/a=1.7}$\/. The number of molecules per channel
length is the same in all cases.}
\end{figure}

At early times before any instability has appeared,
Fig.~\ref{fig:cases} gives head-on snapshots of the cross sectional shape
of the liquid for the two choices of filling and two choices of wetting
considered in this paper. In the two cases of a completely
non-wetting exterior solid shown in Fig~\ref{fig:cases}, the key geometric
parameters for the Stokes equation analysis below are the wetting stripe
width $2\,a$ and the initial liquid height $h_0$\/. 
The case of a narrow stripe corresponds to a width $2\,a=10.6\sigma$ and a 
height-to-half-width ratio $h_0/a=4.0$, while the case of a wide stripe
corresponds to a width $2\,a=17.1\sigma$ with $h_0/a=2.2$\/. 
The maximum system length considered here is $547\,\sigma$ with 128,000 atoms.

First we consider the case of no forcing. If a static liquid ridge of 
uniform cross-sectional shape with contact lines pinned at the chemical
steps (ridges with a mobile contact line are always unstable
\cite{davis80}) is sufficiently long and thick, 
it is expected to be unstable to the formation of bulges, as shown by 
various authors \cite{gau99,davis80,brinkmann02,dtm00}\/. 
This instability is driven by surface tension, 
and is closely related to the well-known Rayleigh-Plateau instability
of a liquid cylinder, which reduces its surface area under long-wavelength
perturbations of its radius, eventually breaking up into spheres. Here, the 
qualification ``thick'' means that the contact angle must be larger than 
$90\degree$ or equivalently, because the liquid cross section in equilibrium 
must be an arc of a circle, the initial height must be greater than half 
the width of the wetting region, i.e., $h_0>a$\/.

\begin{figure}
\includegraphics[width=0.45\linewidth]{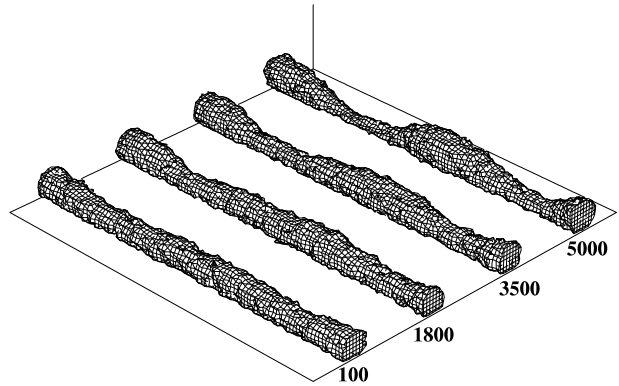} 
\caption{\label{fig:nt0} Instability of a static liquid ridge {with 
${h_0/a=4.0}$} in the nonwetting-narrow case (see
Fig.~\protect\ref{fig:cases}). The sequence runs from 
left to right, showing the median interface position at the times indicated.
The stripe length is $273.6\sigma$ and the vertical axis at the rear of the
figure has a height of $51.3\sigma$. }
\end{figure}

This static instability appears naturally in MD simulations.  In
Fig.~\ref{fig:nt0} we show an example where the initial height was 2.0 
times the stripe width, and the system had 64,000 atoms and half the maximum 
length studied by us.
In this and in subsequent figures, instead of individual atoms we plot the 
median liquid interface, defined as the
surface on which the liquid density averaged over a $50\tau$ interval
falls to half of the value at the
center. A second simulation with initial height $h_0 = a$ was
stable over the same time interval, 
showing continued random fluctuations in shape but no systematic 
evolution. Because the liquid is strongly attracted to the wetting region 
of the substrate, the pearls (with a characteristic wavelength) arising from 
the instability remain
connected by thin liquid films over the duration of the simulation. 
Neither the simulations, nor the linear stability analyses below, can 
reliably predict the ultimate asymptotic state and we cannot determine 
whether the liquid pearls remain connected or not in the true asymptotic
sense $t\to\infty$, i.e., in thermal equilibrium.
Note that the Surface Evolver software used in \cite{gau99,brinkmann02,dtm00} 
does not incorporate the substrate potential as such and cannot address the
issue of adsorbed films of molecular thicknesses. This could be resolved by
density functional techniques combined with a constraint for the liquid
volume; we regard it as likely that the pearls remain connected by thin
liquid bridges. In the simpler case of the pure Rayleigh-Plateau
instability, i.e., in the absence of a substrate, MD simulations similar 
to those described here are in rough agreement with theoretical predictions
of the breakup time; see Refs.~\cite{kb93,k98}\/.

\begin{figure}
\includegraphics[width=0.8\linewidth]{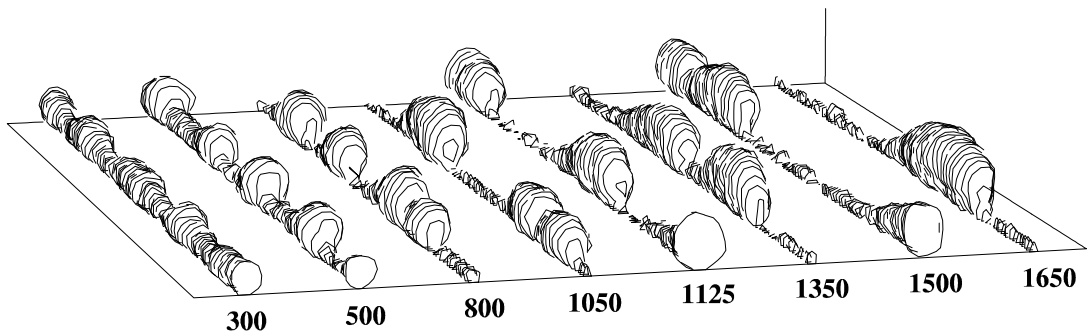}
\caption{\label{fig:nts} Time evolution of the instability of a driven
liquid ridge with the same
aspect ratio and substrate properties as in Fig.~\protect\ref{fig:nt0} 
(i.e., ${h_0/a=4.0}$) but with twice the system size driven at $g=0.01$, 
showing propagating pearls and their merger. The forcing is along the channel
from back to front, and the vertical axis at the
rear of the figure has a height of $51.3\sigma$. }
\end{figure}

\begin{figure}
\includegraphics[width=0.45\linewidth]{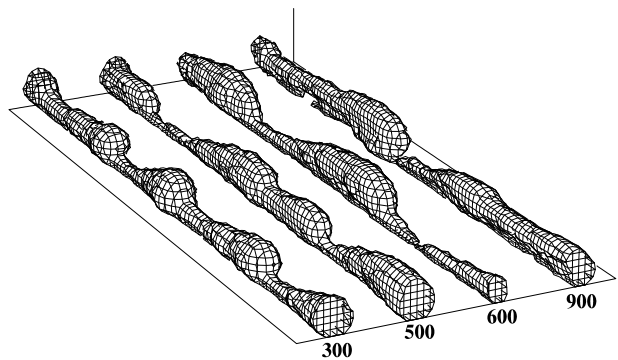}
\caption{\label{fig:ntf} The same system 
as in Fig.~\protect\ref{fig:nts} but driven at a higher acceleration
$g=0.025$\/. Note the difference in time scale for the merging of the
pearls. The vertical axis at the
rear of the figure has a height of $51.3\sigma$.}
\end{figure}

The behavior of the same system when driven with an acceleration 
$g=0.01$ is shown in Fig.~\ref{fig:nts} for the same initial shape. Also
in this case pearls form, with the same initial wavelength as in the 
unforced case, and then propagate downstream with the moving liquid.  
While initially the deviations from the uniform thickness of the ridge are 
periodic and have the same velocity, as they develop to finite amplitude they 
presumably interact nonlinearly and acquire different velocities, and 
eventually merge into pairs until one moving pearl survives. If the length
of the system is half as long, initially only two pearls are present, which 
eventually merge into one; with one quarter of the length, one pearl appears 
and it simply grows and propagates. If the simulation is repeated with $g$ 
increased to 0.025, the same instability scenario is present
(see Fig.~\ref{fig:ntf}) but the structure develops more rapidly, the pearls 
propagate at a higher velocity, and mergers occur sooner.  At still higher 
values of acceleration, in the inertial regime with $\text{Re}>1$, the 
liquid tends to assume very irregular shapes and may even lose contact with
the wetting stripe.

\begin{figure}
\includegraphics[width=0.45\linewidth]{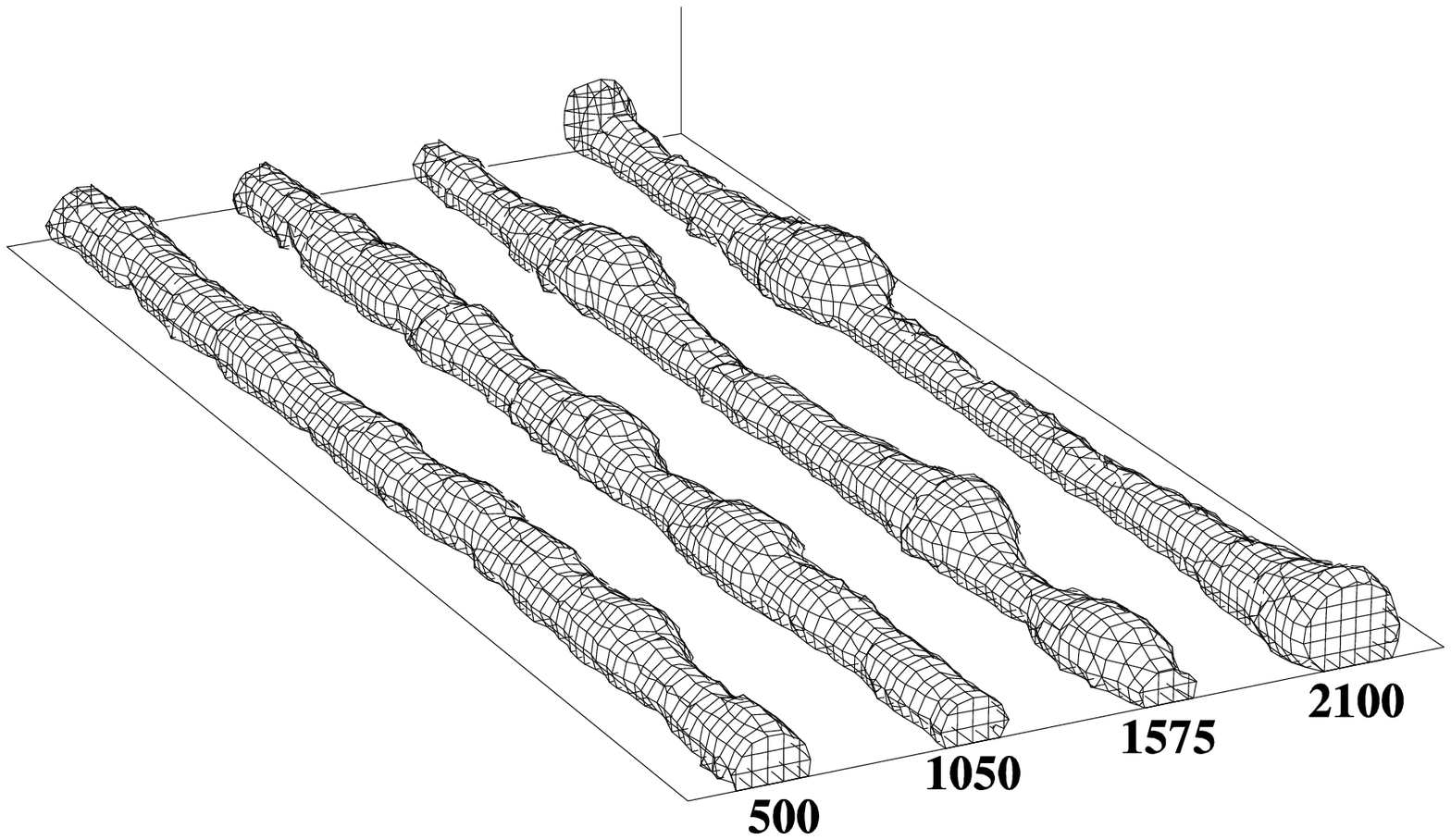}
\includegraphics[width=0.30\linewidth]{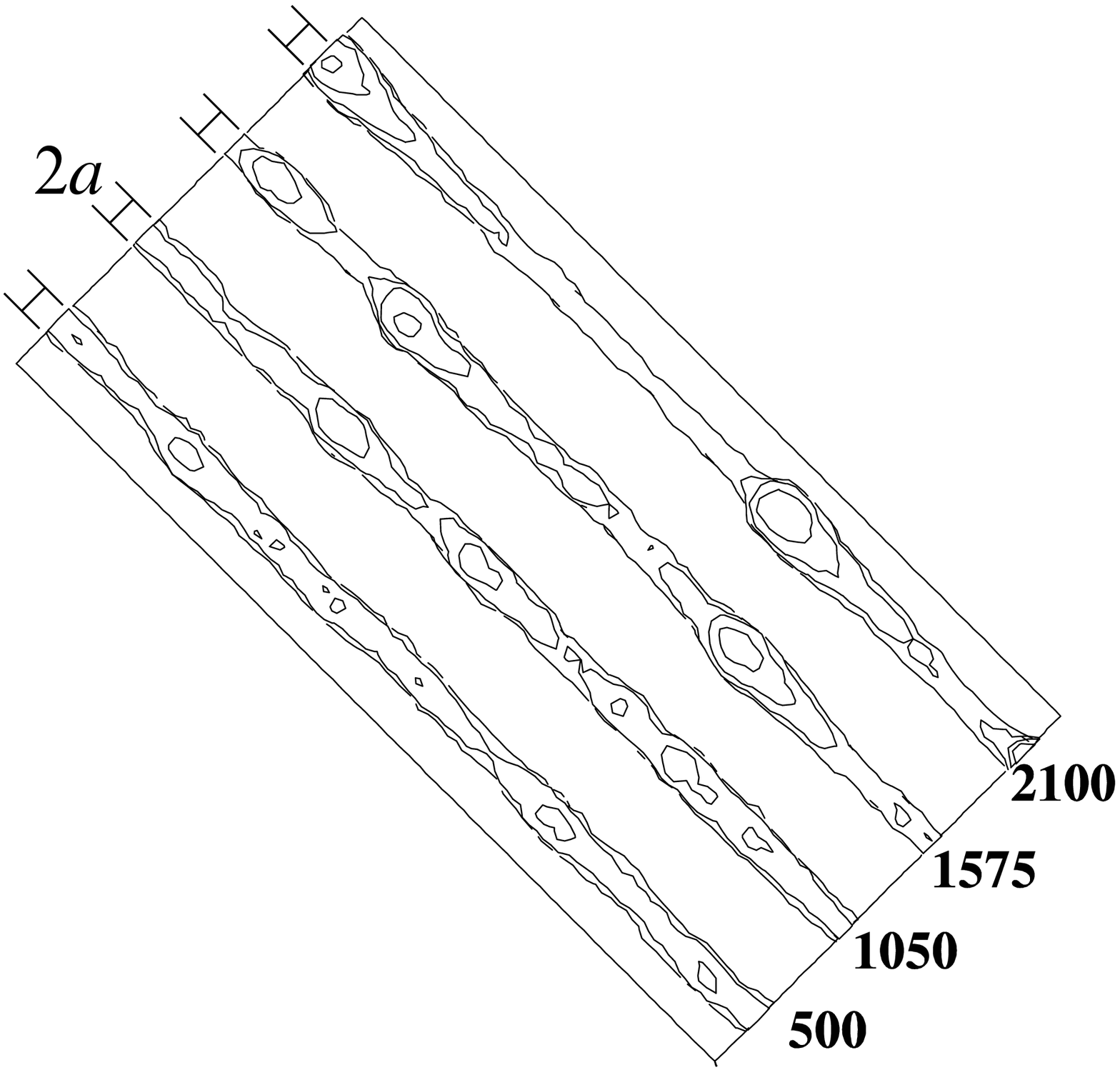}
\caption{\label{fig:pts} Spatio-temporal effects due to a partially-wetting
($c_{LS}=0.75$) exterior solid for a wide wetting stripe ($h_0/a=2.1$) at 
$g=0.01$.  Left: perspective view, right: top view. The vertical axis at the
rear of the left figure has a height of $51.3\sigma$, and the contours in the
right figure are lines of constant height. The scale bars indicate the
width $2\,a$ of the wetting stripe.}
\end{figure}

Somewhat surprisingly, the case that the embedding substrate exhibits
partial wetting shows an identical 
qualitative behavior in terms of pearl formation, motion and merger,
with only numerical differences in growth rate, wavelength, and velocity.  
Generally speaking, in these cases the motion is slower because the liquid 
spills over laterally and tends to be closer to the solid substrate,
because the exterior part of the substrate is more attractive than in
the previous case.
The same number of pearls appear initially but require a longer 
time to develop and to subsequently merge. An example is shown in 
Fig.~\ref{fig:pts} for the case of a wide stripe of width
$2\,a=17.1\,\sigma$ at $g=0.01$, and the principal 
difference from the corresponding results for a non-wetting exterior
is the longer time scale.
The corresponding top views demonstrate that the lateral contact line
formed by 
the liquid on the exterior solid tends to move outward near a pearl, but it
never deviates very far from the edge of the stripe.

A summary of the numerical results for the onset time of the instability, 
its initial wavelength, and the pearl velocity for the cases with a 
non-wetting exterior substrate is given in Table~\ref{tableI} in the rows 
labeled MD, while the other entries are obtained from the stability 
analyses given below. In the table, the ``MD'' numbers assigned to the 
instability are rough estimates obtained by visual inspection. 
The onset 
time $T$ is the time elapsed after the application of the forcing when a
recognizable periodic disturbance (amplitude roughly 10\% of $h_0$) 
appears in a side view of the liquid ridge. The 
wavelength is the average spacing between crests in the pattern, and the 
velocity is obtained from the displacement of the crests of the pearls at 
successive times. The numerical entries for the two analytic calculations 
are the values for the most unstable mode in each case.
The situations involving a partially wetting exterior substrate 
present a rather more difficult stability problem, because in that case
the lateral sides 
of the liquid ridge are mobile instead of being pinned at the stripe edges. 
We have not undertaken this stability analysis.

\begin{table}
\begin{center}
\begin{tabular}{|c||c||c|c||c|c|}
\hline
\multicolumn{2}{|c||}{} & \multicolumn{2}{c||}{wide: $h_0/a=2.2$} &
\multicolumn{2}{c|}{narrow: $h_0/a=4$} \\ \cline{3-6}
\multicolumn{2}{|c||}{} & $g=0.01$ & $g=0.025$ & $g=0.01$ & $g=0.025$ \\
\hline
\hline
               & MD     & 400  & 100-200 & 200  & 100  \\ \cline{2-6}
$T=1/\omega_R$ & LW     & 319  & 319     & 113  & 113  \\ \cline{2-6}
               & NS     & 400  & 229     & 155  & 89   \\
\hline
\hline
               & MD     & 137  & 137     & 137  & 137  \\ \cline{2-6}
$\lambda^*$    & LW     & 168  & 168     & 152  & 152  \\ \cline{2-6}
               & NS     & 107  & 134     & 67   & 83   \\
\hline
\hline
               & MD     & 0.59 & 1.46    & 0.99 & 1.51 \\ \cline{2-6}
$v$            & LW     & 0.46 & 1.15    & 1.06 & 2.65 \\ \cline{2-6}
               & NS     & 0.53 & 0.67    & 0.86 & 1.1  \\
\hline
\end{tabular}
\end{center}
\caption{\label{tableI}
\noindent Comparison of MD simulation, long wavelength (LW) approximation,
and Navier-Stokes (NS) equations stability analysis, for four cases with a
non-wetting exterior substrate. Two stripe widths (wide, $h_0/a=2.2$, and
narrow, $h_0/a=4$) and two accelerations $g$ were considered. $T$ is a
rough estimate of the time at which pearls start to become observable in
the MD simulations, which is compared to $\omega_R^{-1}$, the inverse of
the real part of the instability growth rate. $\lambda^*$ and $v$ are the
distance between the centers of neighboring pearls and their velocity,
respectively.}
\end{table}

Of course, it would be preferable to compare MD and continuum 
results in terms of the time dependence of individual Fourier components 
of the deviation from a uniform axial shape. But in practice two
difficulties prevent this analysis. First, on the atomic scale $\sigma$, 
the simulated liquid ridge exhibits a diffuse interfacial region, so that a
prescription is needed to define a unique liquid-vapor interface.  Second,
independent of how the interface 
is defined, it undergoes capillary wave-like thermal fluctuations
which can obscure the relevant signal. The procedure 
we tried was to choose equally spaced time intervals, to divide the 
liquid ridge into slices of constant-thickness along the flow 
axis, and at each time to count the number of atoms in each bin. Assuming 
that the number of atoms in each bin is proportional to the 
cross-sectional area at that position $z$, and that locally the interface
is an arc of a circle, geometry gives the local height $h(z,t)$ (see, c.f., 
Eq.~(\ref{Aofh}))\/. The power spectrum of $h$ is obtained by a fast Fourier 
transform, from which the time evolution of individual Fourier modes 
follows. If these modes had displayed an exponential growth at early 
times, a precise comparison with a stability analysis would be possible. 
However, we did not observe any systematic behavior of the modal growth 
rates: some increase with time but not exponentially over any appreciable 
time interval, while others simply vary in an irregular manner due to 
thermal fluctuations. In contrast, a recent MD calculation 
of the Rayleigh-Taylor instability in a two-liquid system was able to relate 
the interfacial shape to the growth of distinct Fourier modes \cite{kadau}.  
The presence of the second liquid sharpens the interface and suppresses 
thermal fluctuations, allowing one to make closer contact with the
spectral analysis.

\begin{figure}[h]
\includegraphics[width=0.4\linewidth]{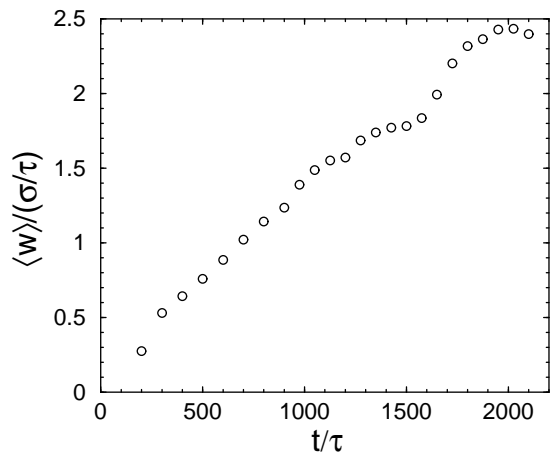}
\caption{\label{fig:vel} Mean axial velocity $\langle w\rangle$ {\it
vs.}\ time for the case of a narrow wetting stripe embedded into a
non-wetting substrate ($h_0/a=4$) as shown in Fig.~\protect\ref{fig:nts}
driven at $g=0.01$\/. }
\end{figure}

Finally, we consider the effects of the pearling instability on liquid 
transport. One effect of pearl formation is to cause a net displacement of
fluid molecules away from the substrate where they are partially bound by
the substrate atoms providing a wetting condition, so that the mean
velocity and the flux should increase as the instability 
grows. From a continuum point of view, the equivalent statement is that 
pearling moves the liquid further away from that part of the boundary on 
which the no-slip condition applies. In fact, as a function of time
we observe a roughly linear increase of the mean \textit{axial} velocity
$\langle w\rangle$ (averaged over the whole liquid ridge) as shown
for one example in Fig.~\ref{fig:vel}\/.  The 
velocities with which the pearls move are comparable to the mean flow 
velocities in each case. In view of the approximate way in which the 
pearl velocities are determined, i.e., 
by monitoring the position of the crests
at regular intervals in time, it is difficult to obtain a more precise
statement. At longer times, the mean velocity oscillates around the maximum 
value
in Fig.~\ref{fig:vel} reflecting shape adjustments of the surviving pearl.
Since in this case $\text{Re} = 0.1$, the roughly linear increase of $\langle
w\rangle$ with time can be attributed to shape changes rather than to inertia.

\section{Stability analysis}
\label{sec:stab}

In this section, we analyze the linear stability of the uniform flow
state on a wetting stripe in the framework of continuum fluid mechanics.
For simplicity we consider stripes of infinite length in order to 
avoid the problem of wavelength selection. Assuming the liquid is 
Newtonian and incompressible, its flow velocity
$\vct{u}=u\vct{e}_r+v\vct{e}_\phi+w\vct{e}_z$ for the geometry shown
in Fig.~\ref{fig:wurst} is governed by the Navier-Stokes equation
\begin{equation}\label{NSeqt}
\frac{\partial \vct{u}}{\partial t}+\vct{u}\!\cdot\!\grad\vct{u}
=-\frac{1}{\rho}\grad p+\frac{\eta}{\rho}\lapl\vct{u}+g\vct{e}_z
\end{equation}
and the incompressibility condition $\grad\cdot\vct{u}=0$, 
where $\rho$ and $p$ are the mass density and the pressure, respectively.
In 
addition, the deformation of the free surface $R(\phi,z,t)$ is described 
by the kinematic condition
\begin{equation}\label{kinematic}
 \frac{\partial R}{\partial t}
+\frac{v}{R}\frac{\partial R}{\partial \phi}
+w\frac{\partial R}{\partial z}=u~,
\end{equation}
which  expresses the fact that (for a non-volatile liquid) a fluid particle
at the interface moves inwards or outwards with the interfacial velocity. 
At the impermeable solid, $\vct{u}$ satisfies the no-slip boundary 
condition (which is consistent with the MD simulations). 
At the free surface, assuming that the liquid is non-volatile,
the appropriate boundary conditions are 
the balance of the normal component of the stress
$\vct{n} \!\cdot\! (\vct{T} \!\cdot\! \vct{n})=-\gamma\kappa$,
and the condition of zero tangential force, 
$\vct{t} \!\cdot\! (\vct{T} \!\cdot\! \vct{n})=0$.
Here $\kappa$ is the local mean curvature of the liquid-vapor
interface, 
$\vct{T}=-p\vct{I}+\eta[(\grad\vct{u})+(\grad\vct{u})^{\rm T}]$ is the 
stress tensor, $\vct{n}$ is the unit normal vector pointing towards the
vapor phase, and $\vct{t}$ with $\vct{n}\cdot\vct{t}=0$ is any 
unit tangent vector on the surface. Since we are considering a non-volatile
liquid, we can assume that the density, the viscosity, and the pressure of
the vapor phase are negligible.

Referring to Fig.~\ref{fig:wurst}, we assume that surface tension and
substrate potential
act -- for a sufficiently large liquid volume -- to
pin the liquid at the edges of the wetting stripe, and to fix the
initial liquid-vapor interface to be a (constant-curvature) section of a 
circular cylinder. 
The contact angle at the edge of the liquid can then take on any value 
between that corresponding to the wetting region on the stripe (0$\degree$) 
and that to the non-wetting one on the exterior (180$\degree$). 
For any given constant initial height $h_0$ and 
acceleration $g$, Eq.~(\ref{NSeqt}) has a steady-state unidirectional flow 
solution $\vct{u}=w_0(r,\phi)\,\vct{e}_z$ which satisfies the 
two-dimensional Poisson equation
\begin{equation}
  {1\over r}\,{\partial\over\partial r}
  \left(r{\partial w_0\over\partial r}\right)  
  +{1\over r^2}{\partial^2 w_0\over\partial \phi^2}
= -\frac{\rho g}{\eta}~.                             \label{Poisson}
\end{equation}
The boundary conditions for $\vct{u}$ reduce to 
$w_0=0$ at the substrate $(|x| \le a, y=0)$ and $\partial 
w_0/\partial n=0$ at the free surface $(r=R, y>0)$. The axially
constrained shape of the initial
liquid-vapor interface can be written as 
\begin{equation}
R(\phi,z,t)\to R_0(\phi)=a\,(b_0\,\sin\phi+\sqrt{1+b_0^2\,\sin^2\phi})
                                                          \label{cylinder}
\end{equation}
with
\begin{equation}
{b_0=\frac{1}{2}\left(\frac{h_0}{a}-\frac{a}{h_0}\right)~.} \label{parab}
\end{equation}
The pressure inside the liquid is $p_0=\gamma/R_c$ where 
$R_c=(h_0^2+a^2)/(2\,h_0)$ is the radius of the truncated cylinder. For
general $h_0/a$, we obtain the base state $w_0$ by solving 
Eq.~(\ref{Poisson}) numerically, using a finite difference discretization.
In the special case in which the liquid occupies exactly half of a
circular cylinder (i.e., $h_0/a=1$), one can obtain an explicit 
Fourier series solution
with the separation ansatz $w_0(r,\phi)=-\rho\,g\, r^2/(4\,\eta) + 
B_0+\sum_{n=1}^\infty
r^n\,\left[A_n\,\sin(n\,\phi)+B_n\,\cos(n\,\phi)\right]$\/. The boundary
conditions and the symmetry of $w_0(r,\phi)$ determine the coefficients 
$A_n$ and $B_n$, and we obtain
\begin{equation}
   w_0(r,\phi)
= \frac{\rho\,g\,a^2}{\eta}\, 
  \left[{1\over 4}\, \left({r\over a}\right)^2(\cos{2\phi}-1)
   -\sum_{n=1,3,5\ldots}\frac{8}{\pi\,n^2\,(n^2-4)} \left({r\over a}\right)^n 
    \sin{n\phi}\right]      \label{ExactSolu},
\end{equation}
which has been discussed in Sec.~\ref{sec:md} and is plotted as the NS
velocity in Fig.~\ref{fig:flow_comp}.

If the liquid starts from rest, the time scale for the steady state 
to develop is controlled by the diffusion of vorticity, which happens 
in the cases simulated above on the time scale
$\tau_D=\rho a^2/\eta \approx 6 \,-\, 16 \,\tau$,
which is much smaller than the characteristic time for the
long wavelength instability to appear.  It is reasonable, therefore, to
study the stability of the flowing liquid ridge by investigating a small
perturbation about the basic unidirectional steady state. Hence, we write 
$\vct{u}=u'\vct{e}_r+v'\vct{e}_\phi+(w_0+w')\vct{e}_z$, $p=p_0+p'$ and 
$R=R_0(\phi)+R'(\phi,z,t)$ and linearize Eqs.~(\ref{NSeqt}),
(\ref{kinematic}), and all the necessary boundary conditions about the
basic state. This leads to a set of linear partial differential equations
(PDE) for the small quantities
$u'$, $v'$, $w'$, $p'$ and $R'$\/. The results are \cite{bat}:
\begin{eqnarray}
     \frac{\partial u'}{\partial t}
&=& -\frac{1}{\rho}\frac{\partial p'}{\partial r}
 +\frac{\eta}{\rho}\left(\lapl u'-\frac{u'}{r^2}
              -\frac{2}{r^2}\frac{\partial v'}{\partial\phi}\right)
 -w_0\frac{\partial u'}{\partial z} \\
     \frac{\partial v'}{\partial t}
&=& -\frac{1}{\rho r}\frac{\partial p'}{\partial\phi}
 +\frac{\eta}{\rho}\left(\lapl v'-\frac{v'}{r^2}
              +\frac{2}{r^2}\frac{\partial u'}{\partial\phi}\right)
 -w_0\frac{\partial v'}{\partial z} \\
     \frac{\partial w'}{\partial t}
&=& -\frac{1}{\rho}\frac{\partial p'}{\partial z}
 +\frac{\eta}{\rho}\,\lapl w'-w_0\frac{\partial w'}{\partial z}
\end{eqnarray}
for the velocity and
\begin{equation}
 \frac{\partial R'}{\partial t}
=u'-\frac{v'}{R_0}\frac{\partial R_0}{\partial\phi}
   -w_0\frac{\partial R'}{\partial z}
\end{equation}
for the kinematic condition in Eq.~(\ref{kinematic}). 
At any time $t$, $p'$ satisfies the pressure equation
\begin{equation}
  \lapl p'
=-2\rho\left(\frac{\partial w_0}{\partial r}\frac{\partial u'}{\partial z}
  +\frac{1}{r}\frac{\partial w_0}{\partial\phi}\frac{\partial v'}{\partial z}
       \right)~,
\end{equation}
which is obtained by taking the divergence of Eq.~(\ref{NSeqt}) and 
using the incompressibility condition.
\begin{figure}
\includegraphics[width=0.4\textwidth]{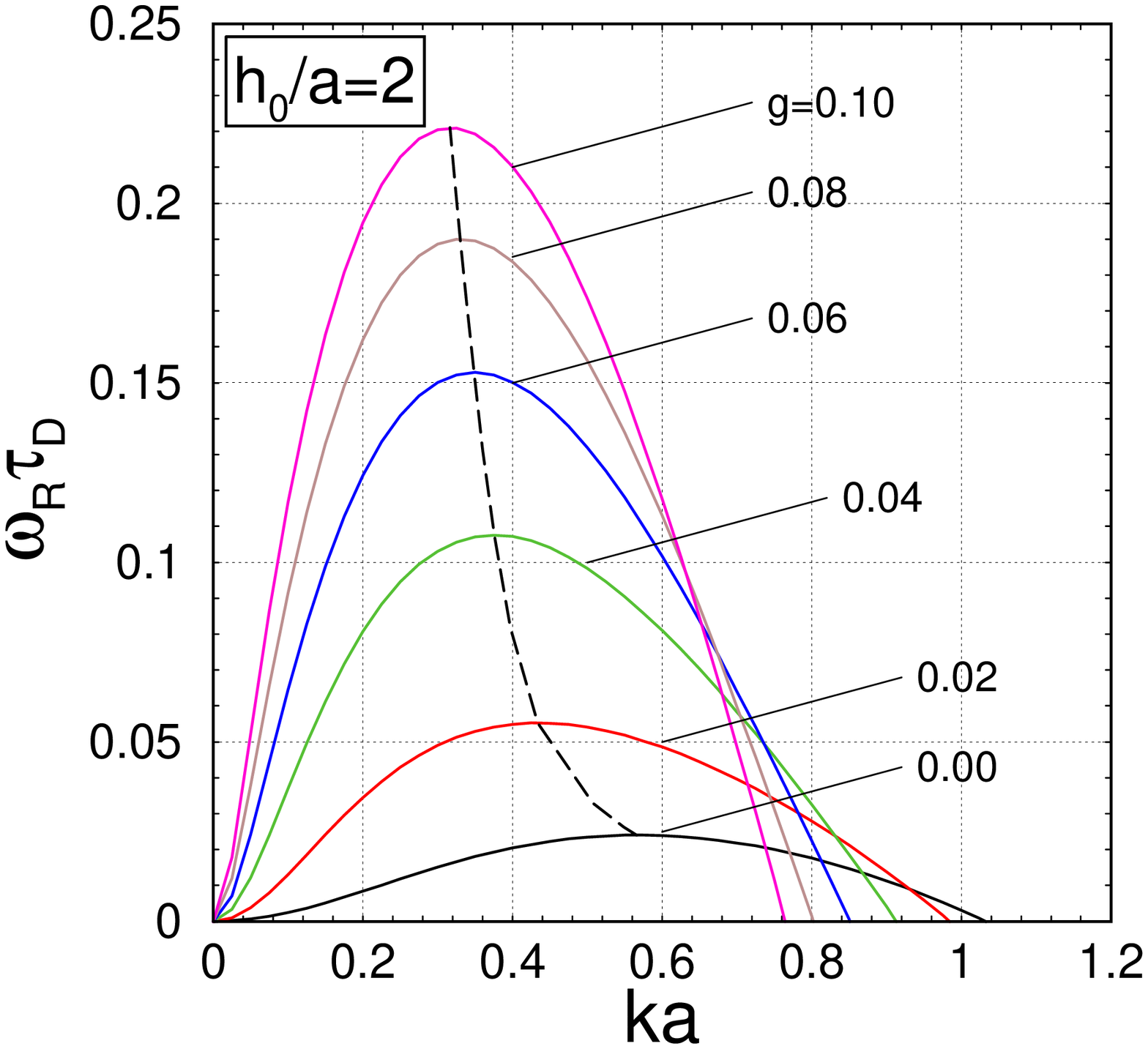}
\includegraphics[width=0.4\textwidth]{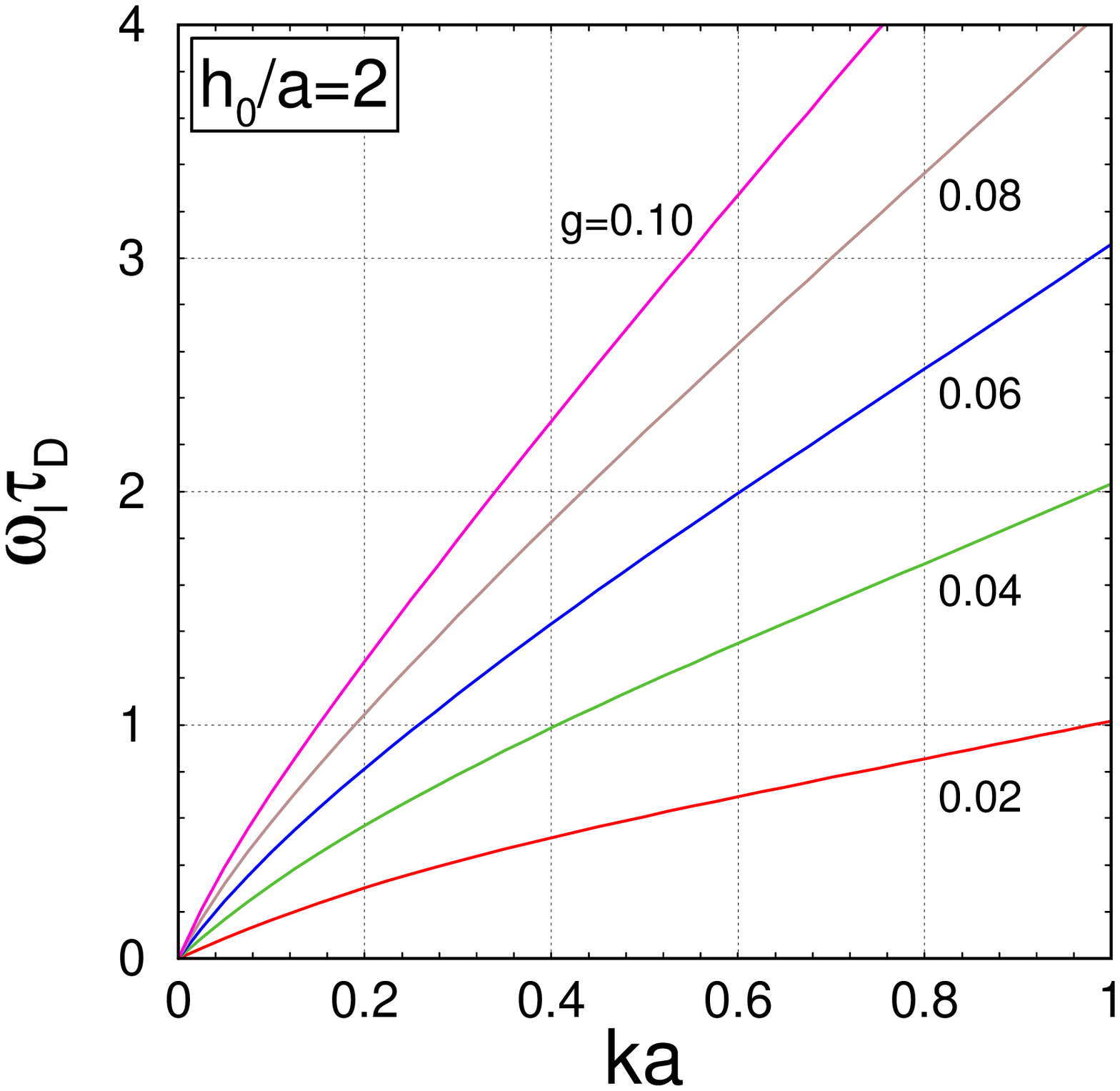}
\caption{\label{fig:LSA} Real (left) and imaginary (right) parts of the
growth rate $\omega(k,1)$ (in units of $\tau_D = \rho\,a^2/\eta$)
of the dominant growing instability modes as function of the wavenumber
$k$ of the perturbation and for $n=1$ (see
Eq.~(\protect\ref{growthrates}))
for a liquid ridge with $h_0/a=2.0$\/. The wavelength of the fastest
growing mode and the corresponding growth rate increase with $g$. The dashed
line marks the loci of these maxima as $g$ varies from $g=0$ to $0.1$
from bottom to top.
The accelerations $g$ are measured in units of $\sigma\tau^{-2}$\/.
Note that $\omega_R( k>k_{\text{max}}(g) )<0$, which implies that
perturbations with shorter wavelengths are linearly stable. Stronger
forcing leads to faster growth of the most unstable perturbation but in the
same time leads to a stabilization of perturbations with shorter
wavelengths. }
\end{figure}

The velocity boundary conditions, upon evaluation at the liquid-vapor
interface position $r=R_0(\phi)+R'(\phi,z,t)$ and 
after linearization, are:
\begin{equation}
 \frac{\partial u'}{\partial z}+\frac{\partial w'}{\partial r}
-\frac{R_{0\phi}}{R_0}
   \left(\frac{\partial v'}{\partial z}
        +\frac{1}{R_0}\frac{\partial w'}{\partial\phi}\right)
-\frac{1}{R_0^2}\frac{\partial w_0}{\partial\phi}
   \left(R_\phi'-\frac{2R_{0\phi}}{R_0}R'\right)
+\left(\frac{\partial^2 w_0}{\partial r^2}
  -\frac{R_{0\phi}}{R_0^2}\frac{\partial^2 w_0}{\partial r\partial\phi}
 \right)R'=0                                   \label{VBC1}
\end{equation}
from $\vct{t}_z \!\cdot\! (\vct{T} \!\cdot\! \vct{n})=0$;
\begin{equation}
\left(1-\frac{R_{0\phi}^2}{R_0^2}\right)
  \left[\frac{\partial v'}{\partial r}-\frac{1}{R_0}
    \left(v'-\frac{\partial u'}{\partial\phi}\right)\right]
+\frac{2R_{0\phi}}{R_0}
 \left(2\frac{\partial u'}{\partial r}+\frac{\partial w'}{\partial z}\right)
-\frac{1}{R_0}\left(1+\frac{R_{0\phi}^2}{R_0^2}\right)
 \frac{\partial w_0}{\partial\phi}R_z'=0     \label{VBC2}
\end{equation}
from $\vct{t}_\phi \!\cdot\! (\vct{T} \!\cdot\! \vct{n})=0$; and
\begin{equation}
 \frac{\partial u'}{\partial r}
=-\frac{1}{R_0}\left(u'+\frac{\partial v'}{\partial\phi}\right)
 -\frac{\partial w'}{\partial z}               \label{VBC3}
\end{equation}
from $\grad\!\cdot\!\vct{u}=0$. The no-slip boundary condition at the
substrate requires $u'=v'=w'=0$ at $\phi=0$ and $\phi=\pi$.

Applying the normal stress condition at the liquid-vapor interface
$r=R_0(\phi)+R'(\phi,z,t)$ gives
\begin{eqnarray}
p'
&\!\!=\!\!& -\frac{\gamma R_{zz}'}{(1+R_{0\phi}^2/R_0^2)^{1/2}} \nonumber \\
& & -\frac{\gamma}{R_0^2(1+R_{0\phi}^2/R_0^2)^{3/2}}
     \left\{R_{\phi\phi}'
    -\left[\frac{R_{0\phi\phi}}{R_0}-\frac{4R_{0\phi}^2}{R_0^2}
    -\left(\frac{1-2R_{0\phi}^2/R_0^2}{1+R_{0\phi}^2/R_0^2}\right)
     \left(1+\frac{2R_{0\phi}^2}{R_0^2}-\frac{R_{0\phi\phi}}{R_0}\right)
     \right]R'\right.                      \nonumber  \\
& & \left.
  -\frac{R_{0\phi}}{R_0}
   \left(
    \frac{1-2R_{0\phi}^2/R_0^2+3R_{0\phi\phi}/R_0}{1+R_{0\phi}^2/R_0^2}
   \right) R_\phi'\right\}                  \nonumber \\
& & +\frac{2\eta}{(1+R_{0\phi}^2/R_0^2)}
    \left[\frac{\partial u'}{\partial r}
     -\frac{R_{0\phi}}{R_0}\left(\frac{\partial v'}{\partial r}
       -\frac{v'}{R_0}+\frac{1}{R_0}\frac{\partial u'}{\partial\phi}\right)
 +\frac{R_{0\phi}^2}{R_0^3}\left(u'+\frac{\partial v'}{\partial\phi}\right)
    \right]             \label{PBC}
\end{eqnarray}
after linearization. In Eqs.~(\ref{VBC1}), (\ref{VBC2}), and (\ref{PBC}),
the subscript $\phi$ denotes differentiation with respect to $\phi$, and 
$r=R_0(\phi)$.

We have solved the linearized equations using the pressure Poisson equation
(PPE) formulation \cite{Gresho87, Johnston02}, which requires boundary
conditions for the pressure $p'$ at the substrate. These conditions can be
obtained from the projections of Eq.~(\ref{NSeqt}) along $\vct{n}$ and
$\vct{t}$, and from the incompressibility condition. To linear order,
they become
\begin{eqnarray}
\frac{1}{r}\frac{\partial p'}{\partial\phi} &=&
\eta\left(\lapl v'+\frac{2}{r^2}\frac{\partial u'}{\partial\phi}\right), \\
\frac{\partial p'}{\partial r} &=&
\eta\left(\lapl u'-\frac{2}{r^2}\frac{\partial v'}{\partial\phi}\right),
\end{eqnarray}
and
\begin{equation}
 \frac{1}{r}\frac{\partial(ru')}{\partial r}
+\frac{1}{r}\frac{\partial v'}{\partial\phi}
+\frac{\partial w'}{\partial z}=0
\end{equation}
at $\phi=0$ or $\pi$. 

We represent
the small quantities in terms of Fourier series expansions, and finally
derive a set of evolution equations for the Fourier components. In this way,
we reduce the three-dimensional problem to a set of two-dimensional linear 
PDEs for each Fourier mode, characterized by a 
wavenumber $k$\/.  We consider 
a small initial sinusoidal perturbation of the liquid-vapor interface of 
the form $R'(\phi,z,t=0)=\epsilon\cos kz\sin\phi$ for 
$|\epsilon| \ll R_0$, and follow its evolution by solving the 
initial value problem numerically. The solution is expanded in a Fourier 
series of the form 
\begin{equation}
  R'(\phi,z,t)
=\sum^\infty_{n=1}\int dk\, A_{nk}(t) e^{-ikz} \sin n\phi\,,
\label{growthrates}
\end{equation}
and the real and imaginary parts of the dominant growth rate
$\omega(k,n) = \omega_R(k,n) + i\,\omega_I(k,n)$
are extracted from the time-dependent amplitude $A_{nk}(t)\propto 
e^{\omega (k,n)\,t}$\/. Modes with $n>1$ are stable. The
functions $\omega(k,1)$ for the 
case $h_0/a=2$ at different values of $g$ are shown in Fig.~\ref{fig:LSA}\/. 
It is evident from Fig.~\ref{fig:LSA} that both the amplitude growth rate 
$\omega_R$ and the phase velocity $\omega_I/k$ increase as $g$ 
increases, which is consistent with the MD simulations. The slight 
change in wavelength of the maximally unstable mode between $g=0.01$ and 
$g=0.025$, which are the values used in the MD simulations, could not be
detected in the simulations for the reasons discussed in Sec.~\ref{sec:md}.
We have carried out the same calculation also for the case $h_0/a=4$ and the
results are qualitatively the same.  For the case $h_0/a=1$, all Fourier 
modes are stable.

\section{Long-wavelength approximation}
\label{sec:lw}

Since the initial liquid configuration is a long ridge with a uniform
cross section, and the structures which develop have, at least initially,
characteristic length scales much larger than the width of the ridge, it is
natural to consider a long-wavelength approximation to the Navier-Stokes
equation. Referring to Fig.~\ref{fig:wurst}, incompressibility implies the
continuity equation
\begin{equation}\label{continuity0}
  \frac{\partial A}{\partial t} = -\frac{\partial Q}{\partial z}
\end{equation}
where $A$ is the local cross-sectional area
\begin{equation}
  A(z,t)=\frac{1}{2}\,\int_0^\pi d\phi\,R^2(\phi,z,t) \label{area}
\end{equation}
and $Q$ is the volumetric flow rate given by
\begin{equation}
Q(z,t)=\int_0^\pi d\phi\int_0^{R(\phi,z,t)} \!\!\!\!dr\,r\,w(r,\phi,z,t)
\end{equation}
with $w(r,\phi,z,t)$ being the axial component of the flow velocity.

If we assume that the local cross section always has a circlular boundary,
the free surface can be described by Eq.~(\ref{cylinder}) with
$b=b(z,t)$ expressed in terms of the local height $h=h(z,t)$ according to 
the same functional relation as between the initial value $b_0$ and the
initial height $h_0$, Eq.~(\ref{parab})\/.  In this case, 
the integral in Eq.~(\ref{area}) yields 
\begin{equation}
A=a^2\, \left[b+(1+b^2)\,\left(\frac{\pi}{2}+\arctan{b}\right)\right]~,
                                      \label{Aofh}
\end{equation}
so that
\begin{equation}
  \frac{\partial A}{\partial t}
= a\, \left(1+\frac{a^2}{h^2}\right)
    \,\left[1+b\, \left(\frac{\pi}{2}+\arctan{b}\right)\right]
   \, \frac{\partial h}{\partial t}~.  \label{dAdt}
\end{equation}

\begin{figure}
\includegraphics[width=0.4\textwidth]{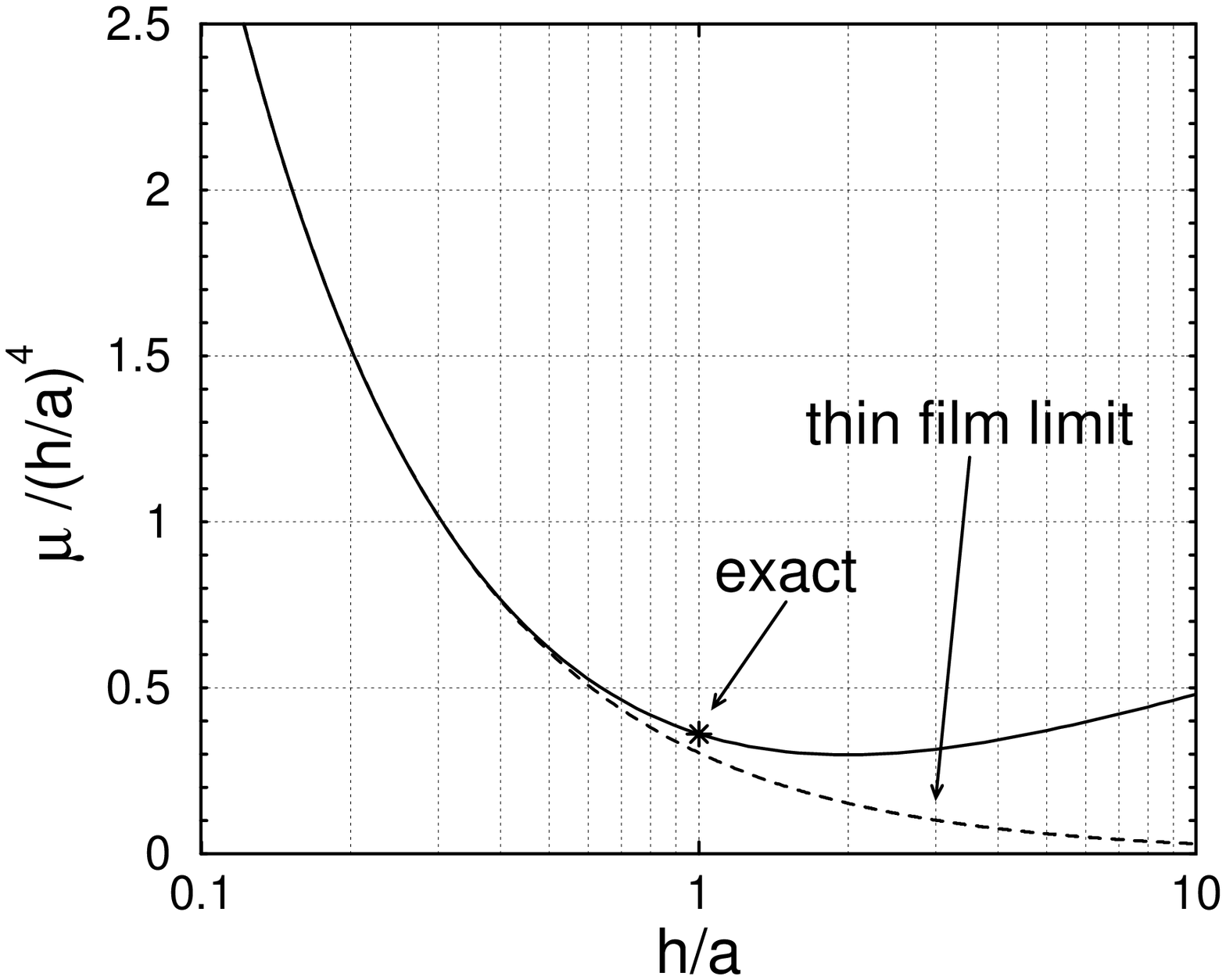}
\includegraphics[width=0.4\textwidth]{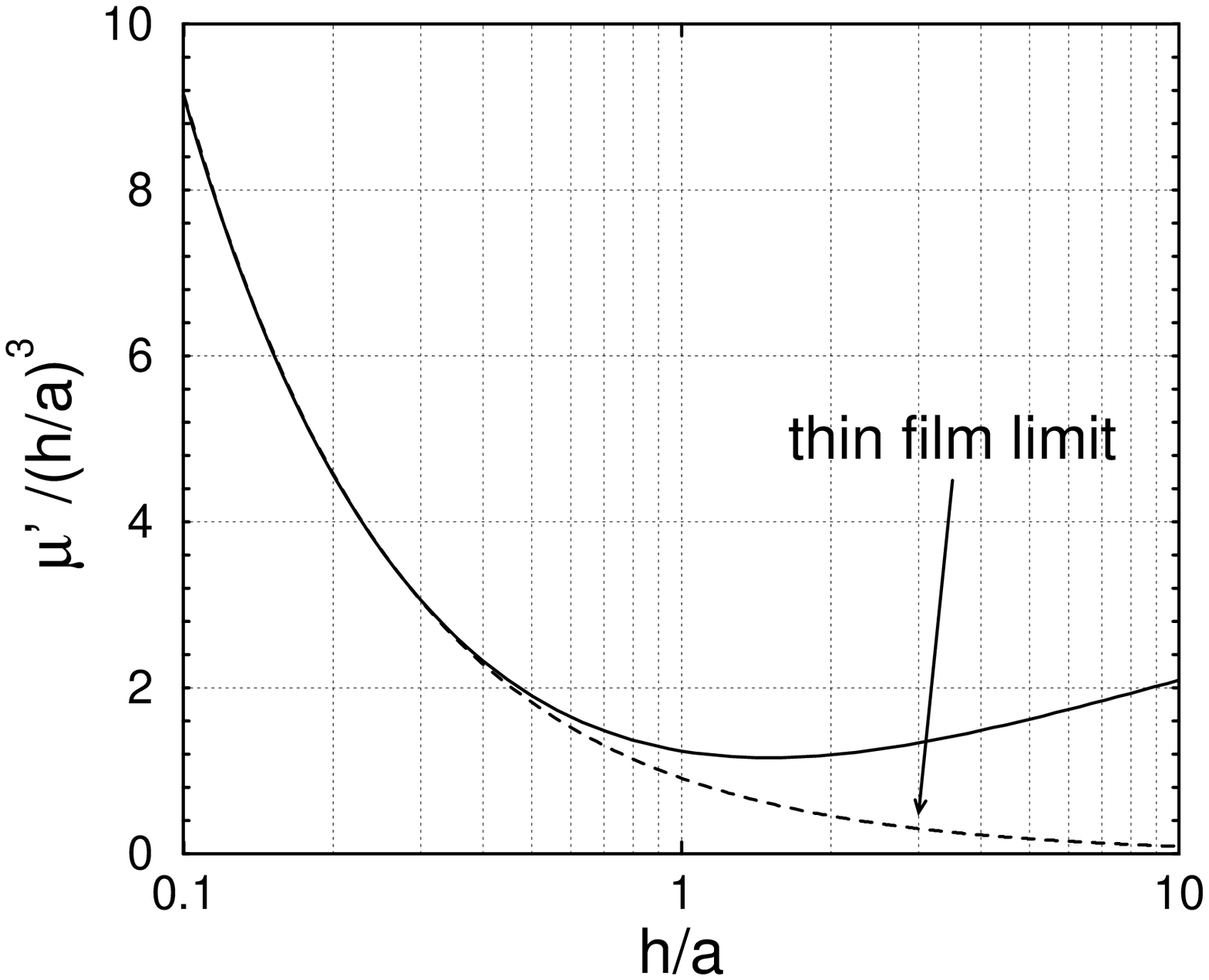}
\caption{\label{fig:lw} The mobility $\mu(h/a)$ as defined in
Eq.~(\protect\ref{flowrate2}) in units of $(h/a)^4$ (left) and its
derivative in units of $(h/a)^3$ (right) from a numerical solution of the
Stokes equation. The dashed lines correspond to the thin film limit $h/a\to
0$.  The value of $\mu(1)$ is known exactly. These units are chosen as
to make the difference between the numerical solutions and the thin film
limit particularly visible. Moreover, it is the combination $a^4\,\mu(h/a)$
which enters into Eq.~(\protect\ref{flowrate2}) for the flux $Q$\/.}
\end{figure}

Instead of performing a systematic long wavelength expansion we 
further assume that $w(r,\phi,z,t)$ can be approximated by the steady 
state axial flow field of a liquid ridge with uniform cross-sectional 
shape given by the local profile $R(\phi,z,t)$.
The flow is driven by the effective forcing 
$(g-\frac{1}{\rho}\frac{\partial p}{\partial z})$, which means that
$w(r,\phi,z,t)$ is given by the solution of Eq.~(\ref{Poisson}) with
$\frac{\rho g}{\eta}$ replaced by 
$(\frac{\rho g}{\eta}-\frac{1}{\eta}\frac{\partial p}{\partial z})$.
This allows for the ansatz
\begin{equation}
 w(r,\phi,z,t)
=w_1(r,\phi;z,t)\left(\frac{\rho g}{\eta}
     -\frac{1}{\eta}{\partial p\over\partial z}\right) \label{AxialVel}
\end{equation}
due to the linearity of the Poisson equation. Here $w_1$ has the
dimension of an area and is proportional to the base state velocity $w_0$
of the previous section because it obeys the same boundary conditions.
Note that although $w_1$ is proportional to the velocity in a uniform infinite
ridge, a function of $r$ and $\phi$, it depends implicitly on $z$ and $t$ 
because of the boundary conditions imposed at $r=R(\phi,z,t)$.
The Laplace pressure $p$ in Eq.~(\ref{AxialVel}) is approximately given by
\begin{equation} 
p = {\gamma\,\kappa} \approx \gamma\,
         \left(\frac{2\,h}{h^2+a^2}-\frac{\partial^2 h}{\partial 
z^2}\right)~,                                      \label{LaplacePress}
\end{equation} 
where the first term $2h/(h^2+a^2)$ is the inverse radius of curvature
of the local circular cross-section set by the local height $h$, and the
second is the (leading order approximation to) the curvature in the
orthogonal plane.  Within this approximation the flux is
\begin{equation}
{Q =  \left[\int_0^\pi d\phi  \int_0^{R_0(\phi)} dr\,r\,
             w_1(r,\phi;z,t) \right]
              \left(\frac{\rho g}{\eta}
                   -\frac{1}{\eta}{\partial p\over\partial z}\right)
  \equiv \frac{a^4}{\eta}\, \mu\bigl(\frac{h}{a}\bigr)
     \, \left(\rho g-{\partial p\over\partial z}\right)~.} \label{flowrate2}
\end{equation}
Since $w_1$ is proportional to $w_0$, the solution for $w_1$ can be
inferred from Sec.~\ref{sec:stab} and in
general it is obtained numerically. The dimensionless ``mobility''
$\mu(h/a)$ (i.e., the proportionality factor between the flux-driving
pressure gradient and the flow rate), 
defined via Eq.~(\ref{flowrate2}), is obtained by numerical 
integration of the solution of Eq.~(\ref{Poisson}). In Sec.~\ref{sec:md} we 
considered the special case $h/a=1$, for which Eq.~(\ref{ExactSolu})
provides the analytical solution, which in turns gives
$\mu(1)=\frac{1}{4}\left\{\frac{1}{\pi}\left[6+7 \zeta(3)\right]-\pi\right\}
=0.362$.
The thin film limit $h/a\to 0$ is an interesting special case as well. 
In this limit, we have a Poiseuille velocity profile locally and
therefore $w_1=w_1(x,y)=\ell(x)y-\frac{1}{2}y^2$ where 
$\ell(x)=h\left[1-(x/a)^2\right]$ is the local film height as a function
of the lateral position $x$.
This leads to $\mu(h/a\to 0)= \frac{32}{105}\,(h/a)^3$.
Figure~\ref{fig:lw} 
shows the mobility $\mu$ and its derivative as functions of $h/a$ 
together with their limiting behaviors for $h/a\to 0$.

\begin{figure}[h]
\includegraphics[width=0.4\textwidth]{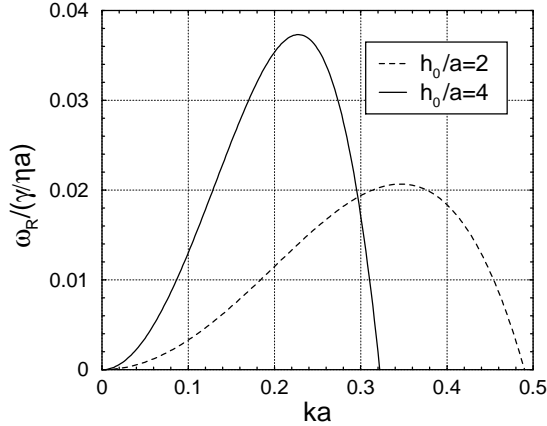} 
\caption{\label{fig:GR_real} 
The real part of the growth rate of unstable modes for two values of 
$h_0/a$ in the long-wavelength approximation (see
Eq.~(\protect\ref{omegaR}))\/. In contrast to the full linear stability
analysis in Sec.~\protect\ref{sec:stab}, the real part does not depend on
$g$\/. $\omega_R$ is negative for $h_0/a<1$\/. }
\end{figure}

With Eqs.~(\ref{dAdt}) and (\ref{flowrate2}), the continuity equation
(\ref{continuity0}) can be re-written as
\begin{equation}
  \left(1+\frac{a^2}{h^2}\right)\,
  \left[1+b\, \left(\frac{\pi}{2}+\arctan{b}\right)\right]
  \frac{\partial h}{\partial t} = \frac{a^3}{\eta}\, 
\left[\mu\bigl(\frac{h}{a}\bigr)\,\frac{\partial^2 p}{\partial z^2}
       -\frac{1}{a}\mu'\bigl(\frac{h}{a}\bigr)
            \left(\rho g-\frac{\partial p}{\partial z}\right)
            \frac{\partial h}{\partial z}\right]~.  \label{continuity} 
\end{equation}
Denoting the local height by $h=h_0+\delta h(z,t)$ and linearizing
Eq.~(\ref{continuity}) around $h_0$ gives
\begin{eqnarray}
  \left(1+\frac{a^2}{h_0^2}\right)
  \left[1+b_0 \left(\frac{\pi}{2}+\arctan{b_0}\right)\right]
  \frac{\partial \delta h}{\partial t} 
& = & -\frac{\gamma\,a^3}{\eta}\mu\bigl(\frac{h_0}{a}\bigr)
   \left[
 \frac{2\,(h_0^2-a^2)}{(h_0^2+a^2)^2}\frac{\partial^2 \delta h}{\partial z^2}
      +\frac{\partial^4 \delta h}{\partial z^4} \right] \\ \nonumber
& &   -\frac{\rho\,g\,a^2}{\eta}\mu'\bigl(\frac{h_0}{a}\bigr)
     \frac{\partial \delta h}{\partial z}
\end{eqnarray}
with $b_0=(h_0/a-a/h_0)/2$. Specializing now to a sinusoidal perturbation
$\delta h \sim e^{\omega\,t-i\,k\,z}$ yields
\begin{equation}
  \left(1+\frac{a^2}{h_0^2}\right)
  \left[1+b_0 \left(\frac{\pi}{2}+\arctan{b_0}\right)\right] \omega = 
\frac{\gamma\,a^3}{\eta}\mu\bigl(\frac{h_0}{a}\bigr)
   \left[\frac{2\,(h_0^2-a^2)}{(h_0^2+a^2)^2}k^2-k^4\right] 
+i\,\frac{\rho\,g\,a^2}{\eta}\mu'\bigl(\frac{h_0}{a}\bigr)k~. 
\end{equation}
From this, one can identify the real and imaginary parts of the growth rate 
$\omega$:
\begin{eqnarray}
    \omega_R
&=& \frac{\gamma}{\eta\,a} \mu\bigl(\frac{h_0}{a}\bigr)
             \left(1+\frac{a^2}{h_0^2}\right)^{-1}
             \left[1+b_0\left(\frac{\pi}{2}+\arctan{b_0}\right)\right]^{-1}
     \left[\frac{2\,(h_0^2/a^2-1)}{(h_0^2/a^2+1)^2}\,(k\,a)^2-(k\,a)^4\right]
                                                   \label{omegaR} \\ 
    \omega_I
&=& \frac{\rho\,g\,a}{\eta} \mu'\bigl(\frac{h_0}{a}\bigr)
             \left(1+\frac{a^2}{h_0^2}\right)^{-1}
     \left[1+b_0\left(\frac{\pi}{2}+\arctan{b_0}\right)\right]^{-1}(k\,a)~.
                                                   \label{omegaI} 
\end{eqnarray}
Graphs of $\omega_R$ for the two cases simulated by MD are shown in 
Fig.~\ref{fig:GR_real}. Equation~(\ref{omegaR}) shows that a long
wavelength
perturbation to a liquid ridge of uniform height is unstable for $h_0>a$ 
(i.e., for contact angles larger than 90$\degree$) and is stable 
otherwise. 
Figure~\ref{fig:GR_real} indicates that for thick ridges the
instability grows more rapidly but only for longer wavelengths.
Equation~(\ref{omegaI}) shows that the phase velocity 
$\omega_I/k$ is proportional to the acceleration $g$, which is 
consistent with the observation in MD that stronger forcing leads to faster 
moving pearls.  However, within this simple approach, the amplitude
growth rate $\omega_R$ is independent of $g$ opposite to what is 
observed in the MD simulations and predicted by the stability analysis in
Sec.~\ref{sec:stab}.
Numerical values for the growth rate, the critical wavelength, and the phase 
velocity are compared with the simulation results in Table I, showing 
at least qualitative agreement.

\section{Summary and Discussion}
\label{sec:disc}

Motivated by possible applications in the field of open 
microfluidic systems we have studied  the flow of a non-volatile 
liquid ridge along a stripe-like
chemical channel, to which the liquid is confined by 
wettability (see Fig.~\ref{fig:wurst}). 
In this context the most salient question concerns stability, i.e.,
the degree to which the liquid remains on the channel and the effects of 
any changes in the shape of the liquid-vapor interface. Our basic tool 
to calculate the flow has been large scale molecular dynamics simulations, 
which directly incorporate the molecular interactions underlying the 
wettability contrast on the flat substrate (see Fig.~\ref{fig:snap} for a
snapshot)\/.  A number of different cases were considered (see
Fig.~\ref{fig:cases}) which varied the amount of liquid present atop the stripe
and the wettability of the exterior region.  In order to understand the
numerical results, we have carried out a systematic linear stability
analysis of the corresponding free boundary problem for 
the incompressible Navier-Stokes equation and find qualitative agreement. 
In addition to the full stability analysis we have presented a much simpler
long-wavelength approximation for channel flow on straight chemical
channels, which is able to capture some but not all of the qualitative
features of the flow.

The basic liquid configuration at rest, i.e., a non-volatile ridge of uniform
cross-sectional shape, is known to be unstable due to a surface tension
driven instability, similar to the Rayleigh-Plateau instability, if 
the contact angle is larger than 90$\degree$ (see Fig.~\ref{fig:nt0}). 
The MD simulations show that when the liquid is driven along the channel,
the instability is enhanced by the driving if it is also present in the
non-driven case (compare the times in Fig.~\ref{fig:nt0} to those in
Figs.~\ref{fig:nts} and Fig.~\ref{fig:ntf}).  However, 
statically stable liquid ridges are not destabilized by the flow. In the 
unstable case a periodic string of pearls appears growing in amplitude 
while propagating along the ridge. At later times, when nonlinear 
effects are manifest, the pearls develop distinct velocities and merge into 
pairs until only one of them survives. Despite substantial changes in
amplitude and shape, the pearls remain confined to the chemical channel up to
rather high accelerations. The instability has a beneficial side effect, in
that the throughput is enhanced as compared to the homogeneously filled case
(see Fig.~\ref{fig:vel})\/. For very high accelerations, however, the 
pearls are heavily deformed and in some cases lose contact with the 
substrate.  A somewhat surprising result is that when the solid exterior to
the wetting stripe is partially wetting rather than completely non-wetting, 
the qualitative behavior is unchanged (see Fig.~\ref{fig:pts}) and the liquid 
remains atop the stripe.

The wavelength of the pearling instability is reasonably well predicted by 
a full linear stability analysis of the Stokes free boundary problem (see
Fig.~\ref{fig:LSA} for the dispersion relations of the most unstable
modes)\/. For the accelerations studied in the MD simulations, the change
in the wavelength of the most unstable mode as compared to the corresponding
static case is too small to be detectable in MD simulations.  However, 
the growth
rate of the most unstable mode increases significantly upon forcing, 
in qualitative agreement with the MD simulations. The results are
summarized in Tab.~\ref{tableI}\/. A precise 
comparison between the MD results and the stability calculations is 
hampered by the thermal fluctuations present in the molecular simulation, 
which are exacerbated by the presence of a free surface which appears as a 
broad interfacial region (see Fig.~\ref{fig:flow_comp})\/. As a result, it is
difficult to identify
from the simulations the individual modes of the stability analysis, and
their growth rates cannot be measured directly. 

The onset of the instability at contact angle $90\degree$ as well as the 
wavelength of the instability are reasonably well predicted by a simple 
long-wavelength expansion of the Stokes free boundary problem. Within this
approximation, the only non-trivial part of the calculation is the numerical
determination of the mobility (see Fig.~\ref{fig:lw})\/.  However, since 
the most unstable wavelength is of the order of the ridge diameter, only 
qualitative results can be expected. In contrast to the MD results as well 
as to the full stability analysis, within this approximation the
maximally unstable wavelength turns out to be independent of the
acceleration (see Fig.~\ref{fig:GR_real} for the real part of the
dispersion relation)\/.

The agreement between the MD simulation results and hydrodynamics
(see Fig.~\ref{fig:flow_comp}) is less 
satisfactory than in other geometries without a free surface. The main
reason seems to be the presence of thermal fluctuations of the
liquid-vapor interface, i.e., capillary wave-like excitations,
which broaden the liquid-vapor interface.  The incorporation of thermal
fluctuations has proven successful in relating MD simulations of a free 
liquid jet to the NS equations \cite{ms00}, and more recently, 
have been included in a hydrodynamic thin film model
\cite{gg05,dms05}. An extension of this technique to liquid ridges with
large contact angles in chemical channels might improve the agreement with
the simulations.

Another possible source of disagreement between the MD simulations and 
the NS results is the neglect of the disjoining pressure
in the hydrodynamic calculations.  The issue arises because large portions
of \textit{thin} liquid films are exposed to the long-ranged substrate
potentials 
and to the absence of liquid molecules outside the films \cite{d88,jni92}, 
and is expected to be important in the nanoscale flows studied here.
In the simple case of a film of nearly constant thickness atop an 
infinite homogeneous substrate, the effect of the disjoining pressure is
captured by adding a contribution $-A_H/h^3$ to the pressure, where $A_H$ is 
the Hamaker constant, which is positive in a completely wetting case.
If we repeat the long-wavelength analysis with this additional term added
to the right hand side of Eq.~(\ref{LaplacePress}), the effect of the
disjoining pressure is to modify the amplitude growth rate to 
\begin{eqnarray}
  \omega_R
&=& \frac{\gamma}{\eta\,a} \mu\bigl(\frac{h_0}{a}\bigr) 
    \left(1+\frac{a^2}{h_0^2}\right)^{-1} 
    \left[1+b_0\left(\frac{\pi}{2}+\arctan{b_0}\right)\right]^{-1}
                                                         \nonumber \\
& & \left\{\left[\frac{2\,(h_0^2/a^2-1)}{(h_0^2/a^2+1)^2}
      -\frac{3A_H}{\gamma a^2}
       \left(\frac{a}{h_0}\right)^4\right]\,(k\,a)^2-(k\,a)^4\right\}~,
                                                         \nonumber
\end{eqnarray}
Hence, the disjoining pressure has a stabilizing effect. It lowers the 
maximum growth rate and 
shifts the most unstable mode towards longer wavelength. In terms of the 
stability criterion, it leads to
$h_0 > a\bigl[1+\frac{3A_H}{2\gamma a^2}
\bigl(1+\frac{a^2}{h_0^2}\bigr)^2\bigr]^{1/2}$
instead of $h_0 > a$. For realistic values of $A_H$ \cite{jni92}, we get 
$3A_H/2\gamma a^2 \sim 10^{-1}$. Such a difference will be masked by the 
uncertainties in MD simulations due to thermal noise and is undetectable
at the present level of accuracy of our results.  This calculation is only 
approximate, since the geometry in this paper is rather more complicated 
than a nearly-uniform film on a homogeneous substrate.
Other authors have extended this disjoining pressure analysis to 
the wedge-shaped geometries arising in drop spreading 
\cite{wedge1,wedge2,BauDi99}, and to
substrates with heterogeneous chemical patterns \cite{d99a,BauDiPa99,d99b}, but 
we feel that dealing with the other effects noted above is more pressing.  
Note that the MD simulations did not include a \textit{long}-ranged
disjoining pressure contribution per se, since the LJ interaction is cut
off. The reason is that in MD simulations of flow at the 10 nm scale, the
principal effect arises from the short-ranged interactions between liquid
and solid atoms, and the long-ranged tail has a negligible effect (see,
e.g., Ref.~\cite{kb00})\/.

For wetting {\em transitions} however, this tail is quite important
\cite{d88}\/. In particular on chemically structured surfaces, the influence
of the laterally inhomogeneous substrate potential on the wetting film
thickness has been discussed in detail \cite{d99a,BauDi99} and on chemical
stripes morphological phase transitions have been predicted
\cite{d99b,BauDiPa99}\/. Recent experiments have confirmed the theoretical
predictions on the structure of wetting films on a chemical stripe
\cite{ocko05}\/. 
However, the thermodynamical ensemble plays a crucial role here. 
There are three types of ensembles which have to be distinguished due to
their distinct characters. If the substrate is exposed to a large vapor
reservoir one deals with a grand canonical ensemble as studied 
theoretically in Refs.~\cite{BauDi99,d99a,BauDiPa99,d99b} and
experimentally in Ref.~\cite{ocko05}\/. For a volatile liquid enclosed into
an isolating container one has a canonical ensemble; the MD simulations
presented here correspond to this case. For a completely non-volatile
liquid the liquid volume in the ridge is strictly conserved; this case 
has been studied in Refs.~\cite{brinkmann05,brinkmann02} and corresponds to
the situation considered for the analytic hydrodynamic analyses in
Sects.~\ref{sec:stab} and \ref{sec:lw}\/. 
The MD simulations are well described by the
non-volatile Stokes dynamics, because the vapor pressure is very low and
the vapor volume is small.
Instabilities, such as the pearling
instability, cannot be observed in a grand canonical setting (in accordance
with the experimental observations in Ref.~\cite{ocko05}), because they only
occur in situations in which the pressure in the liquid ridge decreases
with increasing volume. In this case, in a grand canonical setting,
however, more and more molecules would be drawn out of the reservoir
onto the ridge and it would grow
without limit.  Interestingly, the authors of Ref.~\cite{ocko05} find that
the disjoining pressure does not influence the ridge shape above a ridge
height of approximately 8~nm\/. This finding corroborates our conjecture
given above, that the influence of the substrate potential on the
instability studied in the present paper is small. 

In this work we have considered inertial driving, i.e., a force acting on
every liquid atom in the system independent of its distance from the
substrate surface. Although the physical mechanism for driving the liquid
is qualitatively different, we expect electro-osmotically driven liquids to 
behave very similarly, in particular for low ion concentrations. If in such
systems the Debye layer thickness is on the order of the ridge diameter, the
ions are almost homogeneously distributed inside the ridge and the electrical
force acts equally everywhere. This is in contrast to macroscopic systems,
where the Debye layer is thin as compared to the system size and the
electrical force only acts at the boundary of the system.

Since the pearling instability increases the flow through a chemical 
channel (see Fig.~\ref{fig:vel}), it might be an advantage to operate a
microfluidic system in the unstable regime. Moreover, since the initially
formed pearls have a relatively narrow size distribution, a chemical channel
could be used as a nano-droplet dispenser. For longitudinal driving, the
liquid stays on the chemical stripe even at high accelerations and when the
instability is fully developed. For applications the stability with respect to 
non-longitudinal driving as well as the behavior at junctions and bends 
will be important. Full three-dimensional Stokes flow simulations, 
possibly taking into account the effect of long-ranged interactions and 
thermal fluctuations, should be used in order to interpret the results of 
MD simulations in such complicated geometries.

\begin{acknowledgments}
JK and TSL are supported in part by the NASA Explorations Systems Mission
Directorate. Computational resources were provided by the
Rechenzentrum Garching der Max-Planck-Gesellschaft und des Instituts
f{\"u}r Plasmaphysik.
\end{acknowledgments}

\end{document}